\documentclass[twocolumn]{aastex63}
\usepackage{xspace,amsmath,lineno}

\newcommand{\per}{\ensuremath{^{-1}}\xspace}
\newcommand{\Ha}{H\ensuremath{\alpha}\xspace}

\accepted{January 24, 2022}

\submitjournal{ApJ}

\shorttitle{Subaru/HSC $z$-Broadband Excess  Selection of EELGs at $z<1$}
\shortauthors{Rosenwasser \& Taylor et al.}

\begin{document}

\title{Subaru/HSC $z$-Broadband Excess Selection of Extreme Emission Line Galaxies at $z<1$}

\correspondingauthor{Anthony~J.~Taylor}
\email{ataylor@astro.wisc.edu}

\author{Benjamin~E.~Rosenwasser}
\affiliation{Department of Astronomy, University of Wisconsin--Madison,
475 N. Charter Street, Madison, WI 53706, USA}

\author[0000-0003-1282-7454]{Anthony~J.~Taylor}
\affiliation{Department of Astronomy, University of Wisconsin--Madison,
475 N. Charter Street, Madison, WI 53706, USA}

\author[0000-0002-3306-1606]{Amy~J.~Barger}
\affiliation{Department of Astronomy, University of Wisconsin--Madison,
475 N. Charter Street, Madison, WI 53706, USA}
\affiliation{Department of Physics and Astronomy, University of Hawaii,
2505 Correa Road, Honolulu, HI 96822, USA}
\affiliation{Institute for Astronomy, University of Hawaii, 2680 Woodlawn Drive,
Honolulu, HI 96822, USA}

\author[0000-0002-6319-1575]{Lennox~L.~Cowie}
\affiliation{Institute for Astronomy, University of Hawaii,
2680 Woodlawn Drive, Honolulu, HI 96822, USA}

\author{Esther~M.~Hu}
\affiliation{Institute for Astronomy, University of Hawaii,
2680 Woodlawn Drive, Honolulu, HI 96822, USA}

\author[0000-0002-1706-7370]{Logan~H.~Jones}
\affiliation{Department of Astronomy, University of Wisconsin--Madison,
475 N. Charter Street, Madison, WI 53706, USA}
\affiliation{Space Telescope Science Institute,
3700 San Martin Drive, Baltimore, MD 21218}

\author{Antoinette~Songaila}
\affiliation{Institute for Astronomy, University of Hawaii,
2680 Woodlawn Drive, Honolulu, HI 96822, USA}

\begin{abstract}
We present a search for extreme emission line galaxies (EELGs) at $z<1$ in the COSMOS and North Ecliptic Pole (NEP) fields with imaging from Subaru/Hyper Suprime-Cam (HSC) and a combination of new and existing spectroscopy. We select EELGs on the basis of substantial excess flux in the $z$ broadband, which is sensitive to H$\alpha$ at $0.3\lesssim z \lesssim0.42$ and [OIII]$\lambda$5007 at $0.7\lesssim z \lesssim0.86$. We identify 10,470 galaxies with $z$ excesses in the COSMOS dataset and 91,385 in the NEP field. We cross-reference the COSMOS EELG sample with the zCOSMOS and DEIMOS 10k spectral catalogs, finding 1395 spectroscopic matches. We made an additional 71 (46 unique) spectroscopic measurements with $Y<23$ using the HYDRA multi-object spectrograph on the WIYN 3.5m telescope, and 204 spectroscopic measurements from the DEIMOS spectrograph on the Keck II telescope, providing a total of 1441/10,470 spectroscopic redshifts for the EELG sample in COSMOS ($\sim$14\%). We confirm that 1418  ($\sim$98\%) are H$\alpha$ or [OIII]$\lambda$5007 emitters in the above stated redshift ranges. We also identify 240 redshifted H$\alpha$ and [OIII]$\lambda$5007 emitters in the NEP using spectra taken with WIYN/HYDRA and Keck/DEIMOS. Using broadband selection techniques in $g-r-i$ color space, we distinguish between H$\alpha$ and [OIII]$\lambda$5007 emitters with 98.6\% accuracy. We test our EELG selection by constructing H$\alpha$ and [OIII]$\lambda5007$ luminosity functions and comparing to recent literature results. We conclude that broadband magnitudes from HSC, the Vera C. Rubin Observatory, and other deep optical multi-band surveys can be used to select EELGs in a straightforward manner.
\end{abstract}


\section{Introduction} \label{sec:intro}
Emission line galaxies (ELGs) are direct tracers of star formation throughout cosmic time. Understanding the evolution of their properties, such as their number density, fluxes, and line ratios is a major observational effort. ELGs have remained a topic of great interest since they were first investigated in detail \citep[e.g.,][]{zwicky1964,sargentsearle1970,sargent1970}. ELGs have low stellar masses and metallicities, and they have been used as the basis for searches for the lowest metallicity and/or youngest galaxies and for Lyman continuum leakers and analogs of galaxies that reionized the universe \citep[e.g.,][I. Laseter et al., in prep]{izotov2016,izotov21,naidu21,tang21}. Additionally, ELGs are used as cosmological distance probes using the L-$\sigma$ relation \citep{gonzalezmoran2019}. Their relevance to galaxy evolution is not limited to these examples. 

Decades of research on this heterogeneous class have led to numerous naming conventions including: HII galaxies \citep{sargentsearle1970}, Blue Compact Dwarfs \citep[BCDs;][]{thuan1981}, Ultrastrong ELGs \citep[USELs;][]{kakazu2007}, Green Peas \citep{cardamone2009}, Luminous Compact Galaxies \citep[LCGs;][]{izotov2011}, Extreme ELGs \citep[EELGs;][]{amorin2015} or \citep[XELGs;][]{tanaka21}, Star Forming Dwarf Galaxies \citep[SFDG;][]{grossi2018}, and even Lyman Break Galaxies (LBGs) by the high-redshift ($z\sim3$) IRAC studies \citep[e.g.,][]{magdis08}. The lack of consensus on a naming convention may reflect the rather small size ($\sim$hundreds) of the current spectroscopically confirmed known population. 

This class of galaxy is usually selected using the equivalent widths (EWs) of the \Ha and [OIII]$\lambda$5007 lines. We will refer to these galaxies as EELGs in reference to their extreme EWs. EELGs are easy to identify in grism observations \citep{atek2011,maseda2018,boyett21}, which have confirmed a steep positive evolution in their number density from the local value to higher redshifts \citep{noeske2006}. More recent studies have used slit-based spectroscopy to push these measurements to redshift $z\approx7$ and beyond \citep{endsley21}. The upcoming Near-Infrared (NIR) grism missions \textit{Euclid} and the \textit{Nancy Grace Roman Space Telescope} (\textit{RST}) should detect the full population of these galaxies selected by \Ha emission at $0.7< z < 2.05$ and by [OIII]$\lambda$5007 emission at $1.2 < z < 3$ over tens of thousands of square degrees. The Near-Infrared Imager and Slitless Spectrograph (NIRISS) on the \textit{James Webb Space Telescope} (\textit{JWST}) will extend these sensitivities to $z\approx0.2$ and $z\approx0.6$, respectively, but its comparatively small $2.2'\times2.2'$ field-of-view may limit its usefulness for wide-field EELG surveys.

When extending the large-volume studies of EELGs to lower redshifts where the emission lines of interest are in the optical regime, the NIR grism technique is not applicable. Wide-field optical selections at $z \leq 1$ have typically relied on narrowband imaging, which can be done over large areas \citep{hayashi2018} but in only relatively small volumes due to the intrinsic narrow bandwidth of such filters, which results in a narrow redshift range. Hyper Suprime-Cam \citep[HSC;][]{miyazaki2012} on the Subaru 8.2m telescope, for example, has mapped $\sim$hundreds of square degrees in narrowbands with the largest contiguous area being the North Ecliptic Pole (NEP; A. Taylor et al., in prep). The \textit{Vera C. Rubin Observatory} (\textit{VRO}), will conduct the only deep and wide-area optical survey comparable in sky coverage to the NIR missions, but the telescope will be equipped with 6 broadband ($ugrizY$) and no narrowband filters. Thus, any future selection of \textit{VRO} EELG sources will be limited to broadbands.

Techniques employing wider filters have been used to find EELGs at a variety of redshifts. At $z \leq 0.3$, both intermediate-band \citep{hinojosagoni2016,lumbrerascalle2019} and broadband \citep{cardamone2009,yang2017} filters have been employed. Broadband filters have also been used in the optical to locate intermediate-redshift ($z\sim 0.5$) samples \citep{li2018} and in the NIR to locate high-redshift \citep[$z \geq 1.0$,][]{vanderwel2011,huang2015,tang2019} and very high-redshift \citep[$4 \leq z \leq 8$,][]{smit2015} samples.  Current broadband-selected samples in the optical have used relatively shallow SDSS imaging \citep{cardamone2009,yang2017,li2018}. 

The use of broadband filters when compared to narrowband filters for EELG selection offers several benefits as well as a number of challenges. The most immediate benefit is the larger volume probed by broadband filters, which is typically roughly proportional to the filter width at a fixed central redshift. These larger volumes allow for proportionally larger detected samples of EELGs per unit area on the sky. The cost of this larger volume is an increase in uncertainty in the redshifts of the detected sources, as the detected emission line of interest may generally lie anywhere within the broadband filter passband. Additionally, the sensitivity of a filter to emission line detections at a fixed EW using a narrowband or broadband excess technique is generally inversely proportional to the filter width; thus, narrowband filters have better sensitivity to lower EW sources than broadband filters.

In this work, we select EELGs in the COSMOS and NEP fields using deep HSC imaging in five optical bands ($grizY$). We focus on redshifts $z \leq 1$, which are out of the range of \textit{Euclid} and \textit{RST}. We look for excess flux in the $z$-band relative to the $i$-band and $Y$-band, which is sensitive to \Ha from $0.3 \leq z \leq 0.42$ and [OIII]$\lambda$5007 from $0.7 \leq z \leq 0.86$, and we use the $g-r-i$ broadband color space to separate these two subsamples. We assume $\Omega_{\rm M}$=0.3, $\Omega_{\Lambda}$=0.7, and H$_0$=70~km~s\per Mpc\per throughout. All magnitudes are provided in the AB magnitude system, where an AB magnitude is defined by $m_{AB} = -2.5 \log f_{\nu}-48.60$ and $f_{\nu}$, the flux of the source, is given in units of erg~cm$^{-2}$~s\per~Hz\per. 

\section{Three Filter Broadband Selection}\label{sec_redshift_surveys}
Our COSMOS magnitudes come from the Hyper Suprime-Cam Subaru Strategic Program DR3 (HSC-SSP PDR3; see details in \citealt{aihara21}). We use the COSMOS Deep and UltraDeep datasets, and we select a subregion covering 3.15 deg$^2$. This catalog contains 1,531,398 detected sources.

Our NEP imaging data come from the Hawaii EROsita Ecliptic pole Survey (HEROES) and consist of 34.2 deg$^2$ of deep $grizY$ broadband and NB816, NB921 narrowband imaging. While the HEROES narrowband imaging has been used extensively for selecting ultraluminous Lyman-$\alpha$ emitters at $z>5$ \citep{songaila18,taylor20,taylor21}, in this work we focus on the survey's broadband selection potential.  

The full details on the observations and data reduction for HEROES are given in \cite{songaila18}. Briefly, the HEROES imaging was processed using the Pan-STARRS Image Processing Pipeline (IPP) \citep{magnier20a,magnier20b}. The IPP was also used for source detection. Sources were added to the master catalog with both Kron and forced aperture magnitudes for all 7 filters, if a source were detected at $5\sigma$ in any one of the 7 filters. The full HEROES catalog consists of $\sim$24 million sources, with 91,385 potential $z$-excess EELG candidates.  The $1\sigma$ noise in corrected $2''$ diameter apertures in each broadband are $g$: 27.79, $r$: 27.07, $i$: 27.02, $z$: 26.66, and $Y$: 24.71.

Broadband selection in the optical based on strong [OIII]$\lambda$5007 emission has been performed at low redshift by \cite{cardamone2009} in the $r$-band and by \cite{li2018} in the $i$-band. Here we perform a similar selection using the $z$-band, and we look for an excess relative to the neighboring $i$ and $Y$ bands. This selection traces [OIII]$\lambda$5007 emission in the redshift range $0.7 \lesssim z \lesssim 0.86$, and \Ha emission at $0.3 \lesssim z \lesssim 0.42$. Using a $z$-excess has the advantage of tracing EELGs in two redshift ranges in which they appear in significant numbers, while $r$ and $i$ are sensitive to only [OIII]$\lambda$5007-emitters at lower redshifts.

\begin{figure}[h]
\centering
\includegraphics[angle=0,width=\columnwidth]{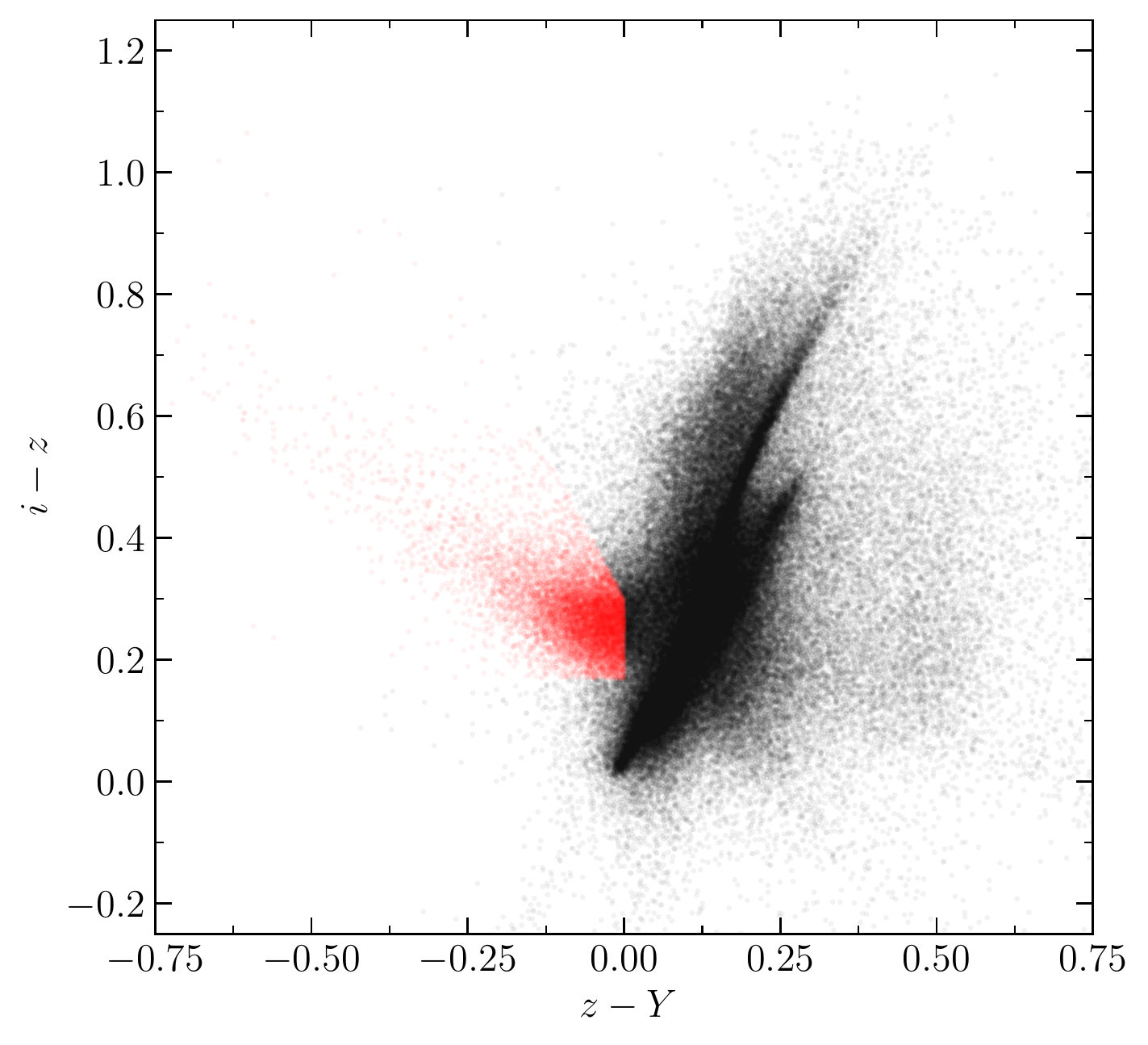}
\includegraphics[angle=0,width=\columnwidth]{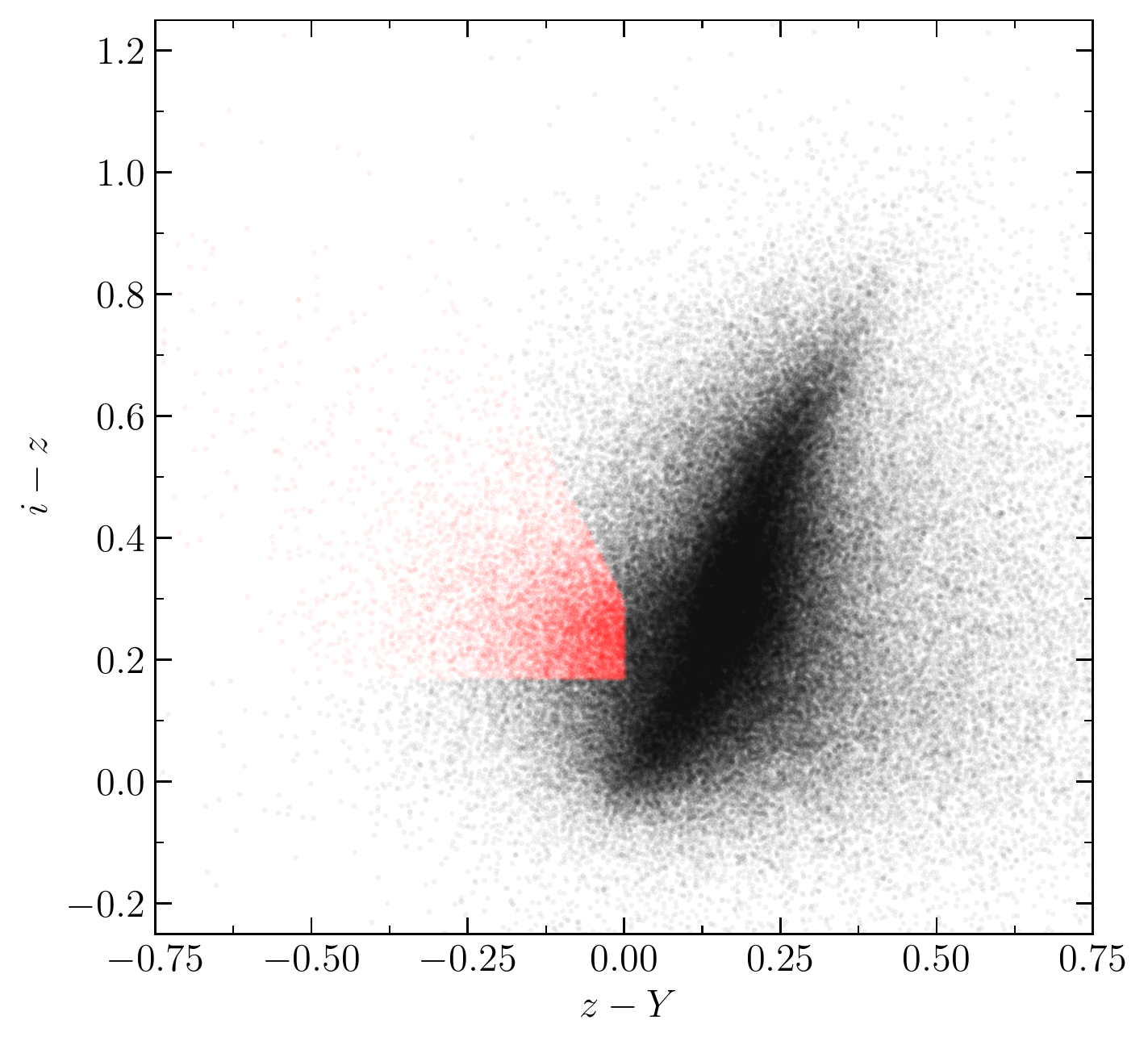}
\caption{\textit{Top:} $i - z$ vs $z - Y$ colors for the COSMOS HSC-SPP catalog. The small black points are the full HSC-SPP catalog with $g,r,i,z<25$ and $Y<23$ while the red points are the $z$-excess sample under study. The $z$-excess galaxies form a substantial outcropping. \\
\textit{Bottom:} Same as above for the HEROES catalog. Note that due to the large $\sim$34 deg$^2$ area, only 11\% of the full catalog sources are plotted to match the COSMOS field size. Due to the superior stacking pipeline and imaging depth in COSMOS HSC-SSP, some finer $i-z-Y$ color space features are less resolved in the HEROES catalog.}
\label{fig:izy}
\end{figure}

In Figure \ref{fig:izy}, we show our color-color selection of EELG candidates. In both fields, we measured $3''$ diameter magnitudes at the positions of all catalog sources in each survey's stacked imaging to ensure consistent magnitude measurements between the two fields. We then selected sources with brightness cuts of $g,r,i,z<25$ and $Y < 23$, corresponding to $\gtrsim 5\sigma$ detections in all five broadbands, to ensure clean detections in $Y$. We next applied principal color cuts of $z - Y < 0.0$ and $i - z > 0.17$. We imposed an additional color cut of $(i-z)+2(z-Y)<0.3$ to reject sources with simultaneous extreme $(i-z)$ colors and moderate $(z-Y)$ colors, which are more indicative of Balmer break galaxies at redshift $z\sim1.4$ or Lyman break galaxies at $z>6$ than of EELGs at redshift $z<1$.

We applied the same cuts to both HEROES and COSMOS, as both fields share similar distributions of sources in the $i-z-Y$ space. In this color-color space, the EELG candidates form a substantial outcropping not seen in other color-color spaces (such as the $r-i-z$ space used by \citealt{li2018}). We also used the HEROES imaging to measure the full-width-half-max (FWHM) of the brightness profile of each source in the $z$-band. We used these data to reject overly compact stellar contaminants and cosmic rays as a function of both FWHM and $z$ filter excess:  FWHM$<0\farcs6-0.1((i+y)/2-z)$. We calibrated this cut using the FWHM data in conjunction with our spectroscopic followup (see Section \ref{sec:specz} below). The combination of all of these cuts produced our $z$-excess sample.

While both fields use HSC imaging, the Deep/UltraDeep COSMOS data are of significantly higher quality than HEROES. This may be due in part to the superior cosmic ray rejection and stacking routines in \texttt{HSCPIPE} \citep{bosch18,bosch19} used to produce the HSC-SSP imaging data and catalogs. As a result, the scatter in the HEROES color-color diagrams is significantly increased compared to the analogous COSMOS diagrams. We are currently remastering HEROES using a combination of new and archival imaging with \texttt{HSCPIPE} for use in future work (A. Taylor et al., in prep). 

\begin{figure}[h]
\centering
\includegraphics[angle=0,width=\columnwidth]{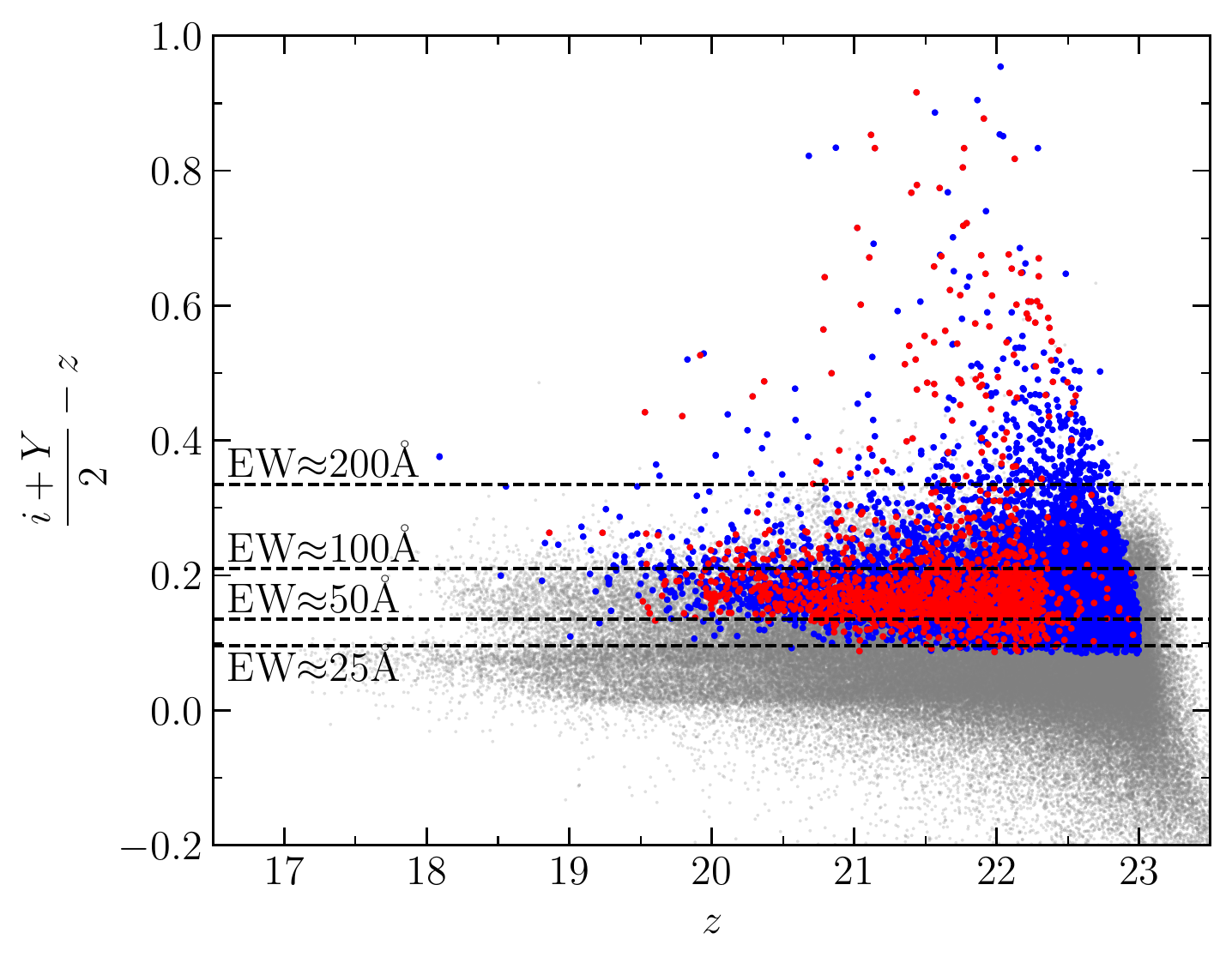}
\includegraphics[angle=0,width=\columnwidth]{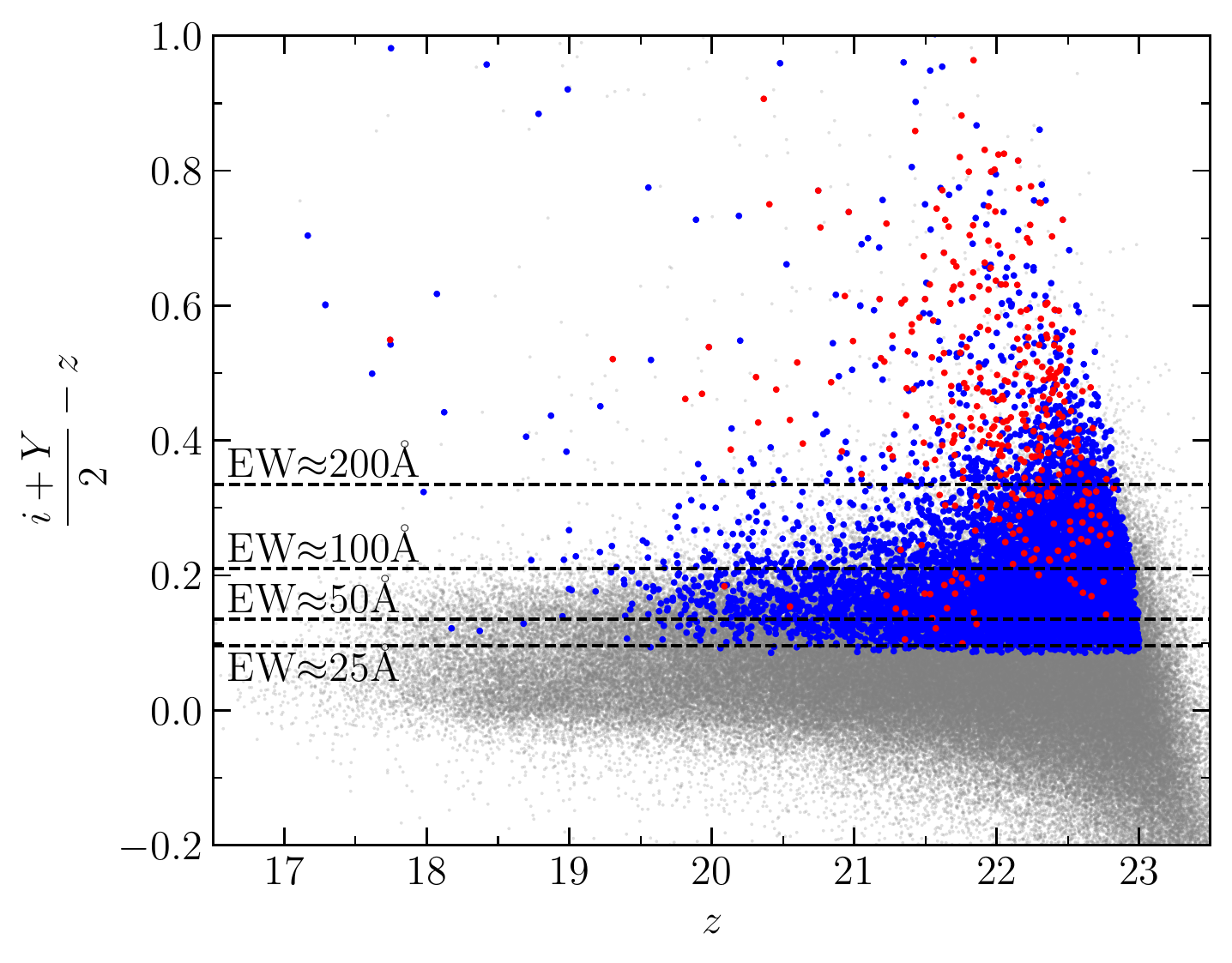}
\caption{$(i + Y)/2-z$ vs $z$ broadband-excess-magnitude diagrams for COSMOS (top) and HEROES (bottom). The small grey points are the full HSC catalog, the blue points are the $z$-excess photometric sample, and the red points indicate galaxies in our sample with observed spectroscopic redshifts. Note that to avoid overcrowding, only 11\% of the full HEROES photometric catalog is plotted. The dashed lines are lines of approximate \Ha or [OIII]$\lambda5007$ observed-frame EWs.}
\label{fig:yzy}
\end{figure}

In Figure \ref{fig:yzy}, we plot $(i + Y)/2-z$ color excess versus $z$ magnitude, with small gray points showing our full COSMOS and HEROES catalogs, blue points showing the photometric $z$-excess samples, and red points marking sources in the photometric z-excess samples with spectroscopic redshifts (discussed below). The $z$-excess samples have a minimum $z$-excess of 0.085, which corresponds to an estimated observed-frame \Ha or [OIII]$\lambda5007$ line EW of $\sim$20~\AA{}. The median $z$-excess is $\sim$0.18, corresponding to an estimated observed-frame \Ha or [OIII]$\lambda5007$ line EW of $\sim$70~\AA{} and thus classifying these as EELGs \citep[e.g.,][]{amorin2015}.  We discuss the calculation and distribution of EWs further in Section~\ref{sec:ew}. We plot lines of constant estimated observed-frame EW in Figure \ref{fig:yzy} as black dashed lines.

Our final $z$-excess samples consist of 10,470 sources in COSMOS and 91,385 in HEROES.

\section{Spectroscopic Redshifts}
\label{sec:specz}
In order to characterize the $z$-excess selection for use with existing and future multi-band catalogs, we calibrated the photometric selection above using newly observed and archival spectroscopic redshifts. First, we matched our COSMOS photometric sample to three spectroscopic surveys in the COSMOS field: zCOSMOS Bright 20k \citep{lilly2009}, DEIMOS 10k \citep{hasinger18}, and our WIYN targeted samples (described below). We obtained 1245 matches from zCOSMOS, 204 from DEIMOS, and 71 from WIYN, for a total of 1520 redshifts (1441 unique), or 13.8\% spectroscopic completeness. 

The spectroscopic coverage provided by zCOSMOS only probes to $i\sim 22.5$, much shallower than the HSC imaging. In order to probe the fainter magnitudes probed by the COSMOS HSC imaging, we performed additional spectroscopic follow-up for a small number of COSMOS sources with $Y\lesssim$23 using the multi-object spectrograph HYDRA on the 3.5m WIYN telescope at Kitt Peak National Observatory. The observations were made in a series of runs in January, February, and March 2016 that targeted both broadband and narrowband excess candidates. We used HYDRA's `red' fiber bundle which is optimized for $5000 - 9500$~\AA{} throughput and contains $\sim$80 fibers with $2\farcs0$ diameters and a positional accuracy of $0\farcs3$. About 10 fibers per configuration were placed in random sky locations and combined to produce an average sky spectrum. We configured the spectrograph using the 316@7.0 grating and GG-420 filter, providing a spectral resolution of 2.6~\AA{} pixel\per . Second-order contamination did not appear to affect our redshift measurements. Calibrations were done for each configuration with a CuAr lamp. The reduction was carried out with standard IRAF routines for dark and bias subtraction and flat fielding, and the IRAF task \texttt{dohydra} was used for sky subtraction and dispersion calculation. Redshifts were measured by eye on the basis of the strong emission lines and were later refined through spectral fitting, using the by eye redshifts as initial guesses for the fitter (see Section \ref{sec:lineflux}). Based on the 2.6~\AA{} pixel\per spectral resolution, we estimate our $1\sigma$ redshift errors at $\pm0.002$. We obtained redshifts for 71 total sources from the photometric sample, 46 of which had not been previously measured in other surveys. 

\begin{figure}[h]
\centering
\includegraphics[angle=0,width=\columnwidth]{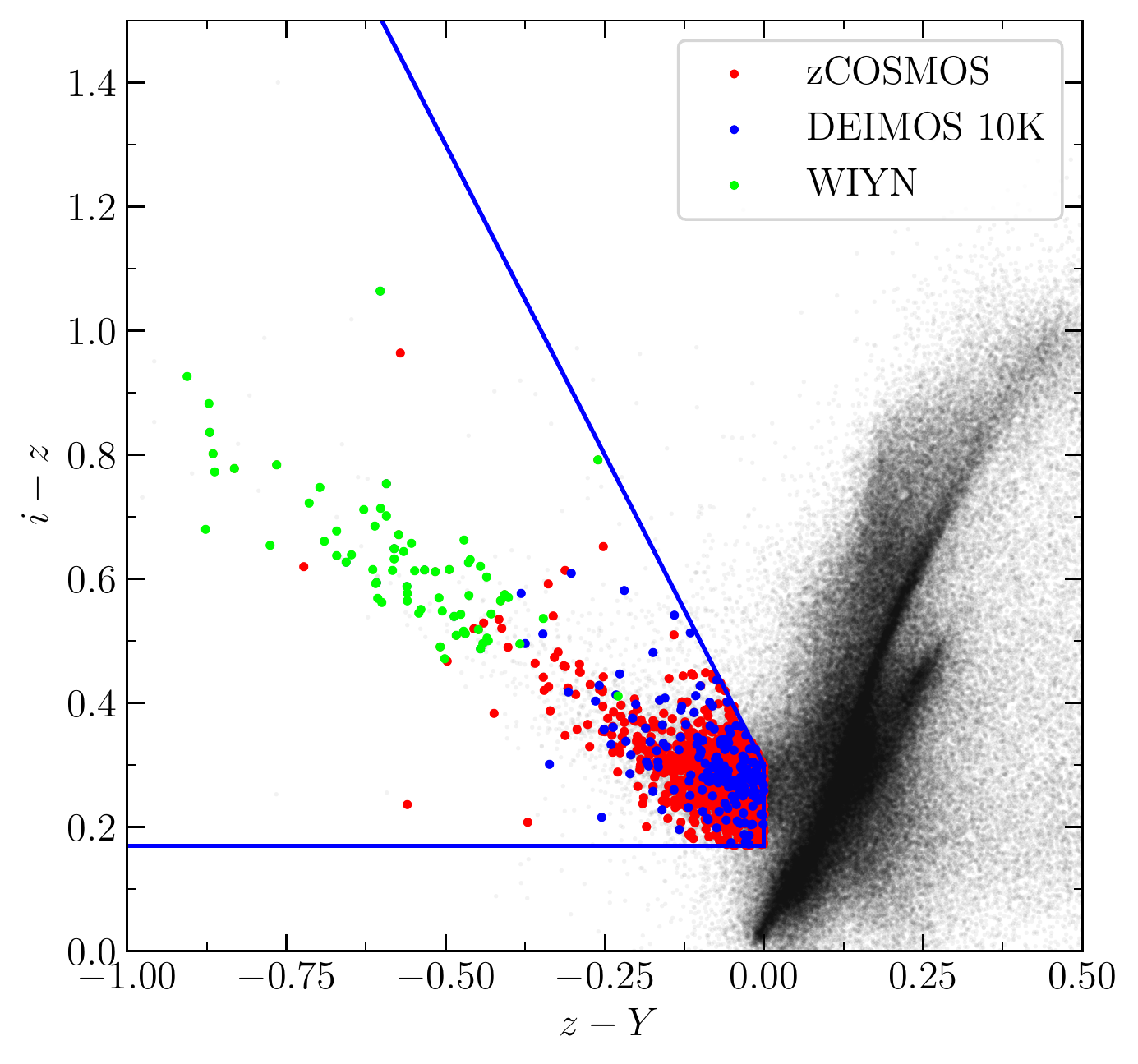}
\includegraphics[angle=0,width=\columnwidth]{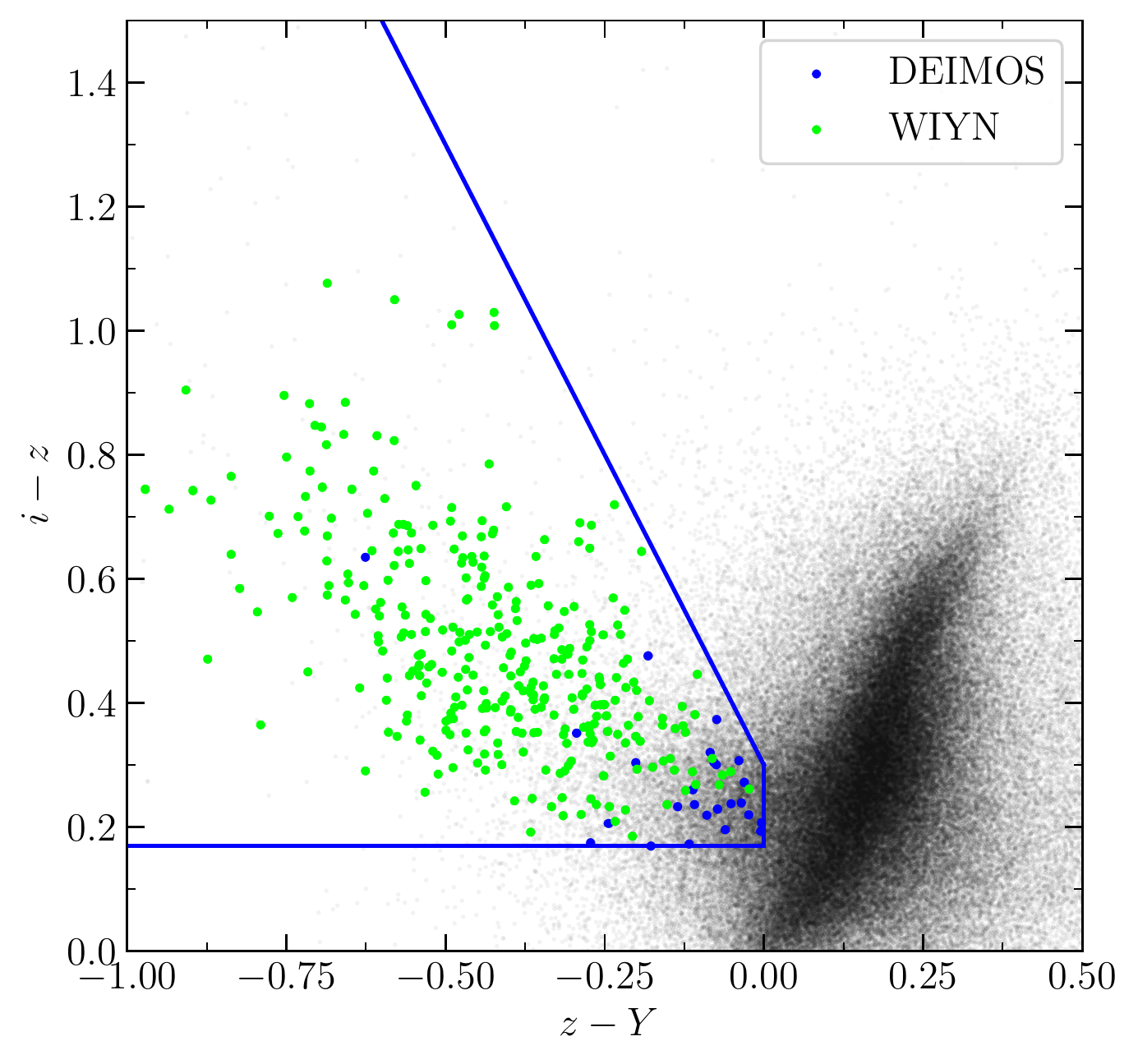}
\caption{The spectroscopic samples in COSMOS (top) and HEROES (bottom) plotted in $i-z-Y$ color-color space. Fine black points show the entire  photometric catalogs, while red points show spectroscopic matches from zCOSMOS, blue points from DEIMOS, and green points from HYDRA. The $i-z-Y$ color selection cuts are shown by the blue borders.} 
\label{fig:cosmosspecs}
\end{figure}

We plot all spectroscopically observed sources in $i-z-Y$ color space in Figure \ref{fig:cosmosspecs} (top). Note that the sources that we observed with WIYN tended to have larger $z$-excesses than the more generally selected targets of zCOSMOS and DEIMOS 10K. This selection bias was by design, both to better sample the $i-z-Y$ color-color space and to search for high EW EELGs. 

After verifying the success of the $i-z-Y$ selection in COSMOS, we used the same HYDRA configuration to observe sources in the spectroscopically unexplored NEP during runs in June 2017, October 2017, May 2018, and September 2018. In total, we observed 331 candidates and obtained redshifts for 240 (73\%) of them. Many of the remaining 91 sources showed blank spectra with no discernible emission lines or features. Upon further inspection, around a third of these sources also showed nearby noise and neighboring sources in the HSC imaging. 

We supplemented our catalog of WIYN/HYDRA spectra with sources from various Keck/DEIMOS runs that had narrowband selected Ly$\alpha$ emitters at $z>5$ as the primary targets \citep{hu2016,songaila18,taylor20,taylor21}. This catalog contains spectroscopic redshifts for 1191 objects at $z<6.6$.  Of these, 33 match candidates in the $z$-excess sample for a total of 364 (357 unique) redshifts in HEROES. 

Due to the large area of HEROES, this represents only a small fraction 357/91,385 (0.39\%) of the identified $z$-excess candidates. However, the total (HYDRA+DEIMOS) sample of 357 spectroscopically identified candidates is sufficient as a sample of HEROES EELGs and as further verification that the $i-z-Y$ selection methods work. We plot this sample of spectroscopic sources in Figure \ref{fig:cosmosspecs} (bottom). Note that as in COSMOS, the HYDRA spectra sample a much more extreme color-color space than the narrowband selected DEIMOS spectra. This is again by design, as we were interested in searching for particularly strong EELG sources.

\begin{figure}[h]
\centering
\includegraphics[angle=0,width=\columnwidth]{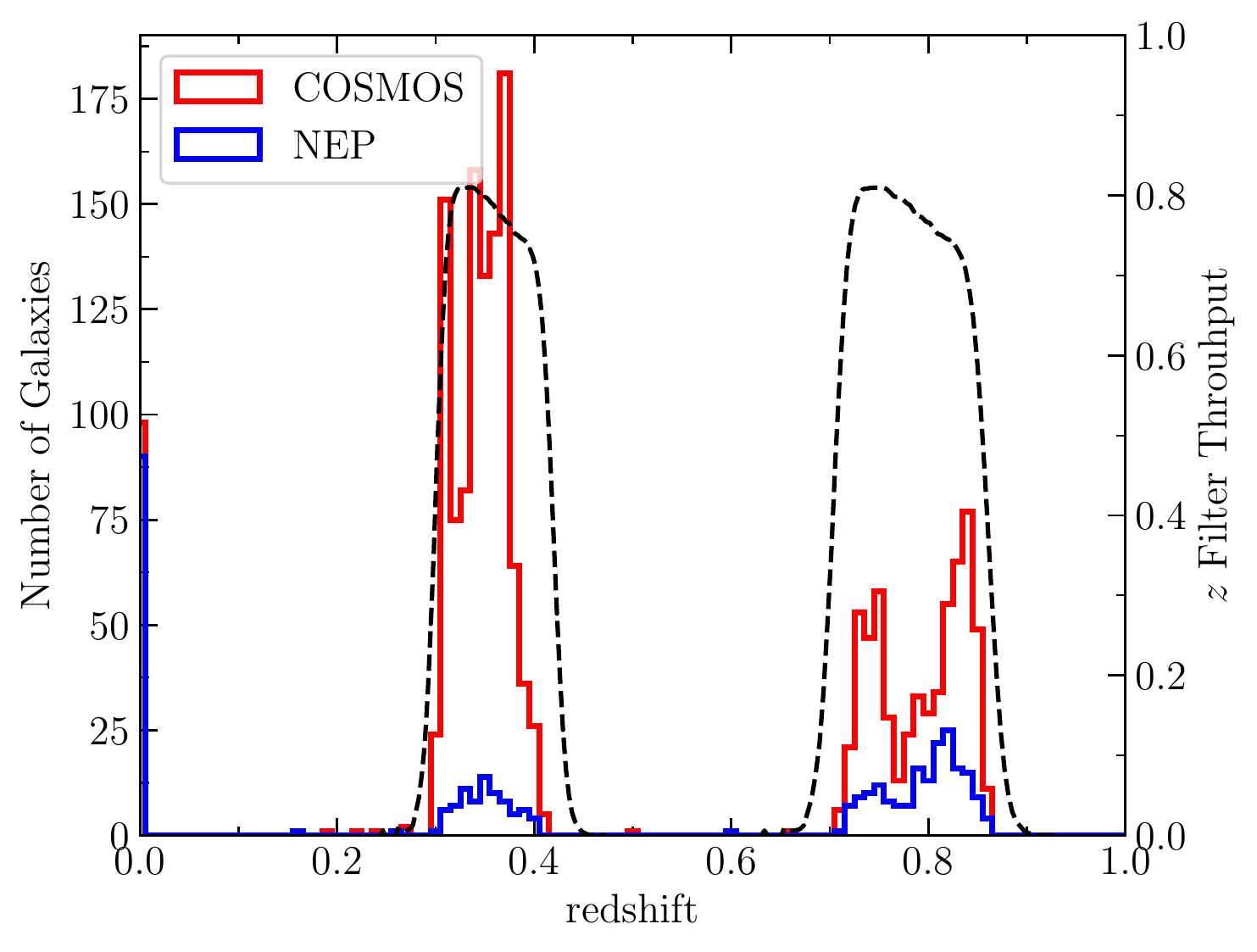}
\caption{Stacked redshift distribution of the 1441 $z$-excess objects in COSMOS (red) and 357 in HEROES (blue) that were spectroscopically observed. The dashed black profiles correspond to the redshifts where \Ha and [OIII]$\lambda$5007 are within the $z$ filter passband. Sources with spectroscopic followup but no discernible spectral features and no evident redshift are assigned a redshift of $z=0$ in this figure.} 
\label{fig:red_hist}
\end{figure}

In Figure \ref{fig:red_hist}, we show the distribution of redshifts for the spectroscopically observed $z$-excess objects in both fields. The dashed black profiles correspond to the $z$-filter transmission at the redshifts where \Ha and [OIII]$\lambda$5007 are within the $z$ passband. The redshift distribution is highly peaked in the two redshift ranges defined by \Ha and [OIII]$\lambda$5007, $0.30 \lesssim z \lesssim 0.42$ and $0.7 \lesssim z \lesssim 0.86$. The bimodal shape of the distribution of [OIII] emitters in COSMOS results from a combination of effects. \cite{hasinger18} noted this structure in the full DEIMOS 10K Bright survey and identified a protocluster and several filaments that caused the over-density of sources at redshifts $z\approx0.73$ and $z\approx0.84$. Coincidentally, the peak at redshift $z\approx0.84$ also corresponds to [OIII]$\lambda$5007 entering a NB921 narrowband. As some of our WIYN/HYDRA objects were selected with narrowband imaging for use in other projects, these sources further accentuate the bimodal structure. The narrowband selection of DEIMOS sources in the NEP causes the same effect to a lesser extent. 

1679/1789 galaxies (94\%) fall in the two redshift ranges defined by redshifted \Ha and [OIII]$\lambda$5007 emission, while the remaining 110 are outside of these ranges, of which 107 showed blank or indeterminate spectra.  We thus confirm that the selection is primarily sensitive to \Ha-emitters (HAEs) and [OIII]$\lambda$5007-emitters (O3Es) at $0.3 \lesssim z \lesssim 0.42$ and $0.7 \lesssim z \lesssim 0.86$, respectively. Hereafter, when referring to HAEs or O3Es, we also imply that each species lies in these redshift bounds, and that [OIII] refers to [OIII]$\lambda5007$ unless otherwise stated. Our 94\% overall sample purity is very comparable to the purities seen in narrowband surveys at similar redshifts, e.g., 95\% for \Ha at  $z=0.84$ \citep{sobral13}, 93.7\% for \Ha at $z=0.47$, and 90.4\% for [OIII] at $z=0.93$ \citep{LAGER20}. We summarize our spectroscopically confirmed HAEs and O3Es in Table~\ref{tab:zspec}.

\begin{deluxetable*}{cccccccccc}[ht]
\renewcommand\baselinestretch{1.0}
\tablewidth{0pt}
\tablecaption{Selected Spectroscopic Redshifts}
\label{tab:zspec}
\tablehead{Field & RA (J2000)& Dec (J2000)& $g$ & $r$ & $i$ & $z$ & $Y$ & Redshift & Instrument}
\startdata
COSMOS & 149.27763 & 2.04143 & 22.56 & 22.28 & 22.12 & 21.44 & 22.31 & 0.717 & HYDRA \\
COSMOS & 149.29378 & 2.27055 & 23.79 & 23.46 & 23.04 & 22.49 & 22.92 & 0.847 & HYDRA \\
COSMOS & 149.35527 & 2.55651 & 23.50 & 23.26 & 22.93 & 22.29 & 22.94 & 0.820 & HYDRA \\
COSMOS & 149.50330 & 1.78595 & 23.44 & 23.17 & 22.86 & 22.39 & 22.89 & 0.789 & HYDRA \\
COSMOS & 149.52539 & 1.94894 & 22.43 & 21.94 & 22.08 & 21.51 & 21.91 & 0.371 & HYDRA \\
COSMOS & 149.63052 & 2.38696 & 23.06 & 22.54 & 22.81 & 22.27 & 22.75 & 0.339 & HYDRA \\
COSMOS & 149.70924 & 1.44678 & 23.04 & 22.82 & 22.58 & 21.96 & 22.58 & 0.788 & HYDRA \\
COSMOS & 149.72950 & 2.49113 & 23.04 & 22.99 & 22.90 & 22.12 & 22.99 & 0.832 & HYDRA \\
COSMOS & 149.77736 & 1.44253 & 22.44 & 22.22 & 21.95 & 21.38 & 21.89 & 0.784 & HYDRA \\
COSMOS & 149.81691 & 2.33237 & 23.37 & 22.99 & 22.62 & 22.06 & 22.60 & 0.761 & HYDRA \\
\nodata& \nodata& \nodata& \nodata& \nodata& \nodata& \nodata& \nodata& \nodata& \nodata\\
NEP & 273.68683 & 65.39443 & 22.93 & 22.89 & 22.66 & 22.03 & 22.65 & 0.846 & DEIMOS \\
NEP & 273.70792 & 67.62094 & 23.18 & 22.68 & 22.87 & 22.30 & 22.54 & 0.312 & HYDRA \\
NEP & 273.95770 & 67.06726 & 23.20 & 22.62 & 22.40 & 22.11 & 22.25 & 0.352 & HYDRA \\
NEP & 274.22275 & 67.21378 & 23.04 & 22.74 & 22.44 & 21.97 & 22.29 & 0.744 & HYDRA \\
NEP & 274.25754 & 68.67712 & 22.59 & 22.19 & 22.10 & 21.70 & 22.12 & 0.366 & HYDRA \\
NEP & 274.26724 & 68.24654 & 22.87 & 22.74 & 22.29 & 21.96 & 22.23 & 0.837 & HYDRA \\
NEP & 274.30870 & 67.60469 & 23.03 & 22.35 & 22.86 & 22.22 & 22.70 & 0.360 & HYDRA \\
NEP & 274.32544 & 68.20570 & 22.92 & 22.87 & 22.84 & 21.96 & 22.67 & 0.780 & HYDRA \\
NEP & 274.42224 & 67.83610 & 22.81 & 22.66 & 22.09 & 21.79 & 21.86 & 0.815 & DEIMOS \\
NEP & 274.42390 & 67.76432 & 22.48 & 22.02 & 21.79 & 21.57 & 21.59 & 0.313 & DEIMOS \\
\enddata
\tablecomments{This table is available in its entirety in a machine readable format.}
\end{deluxetable*}

\section{Five Filter Broadband Selection}

\begin{figure*}[h]
\centering
\includegraphics[angle=0,width=\columnwidth]{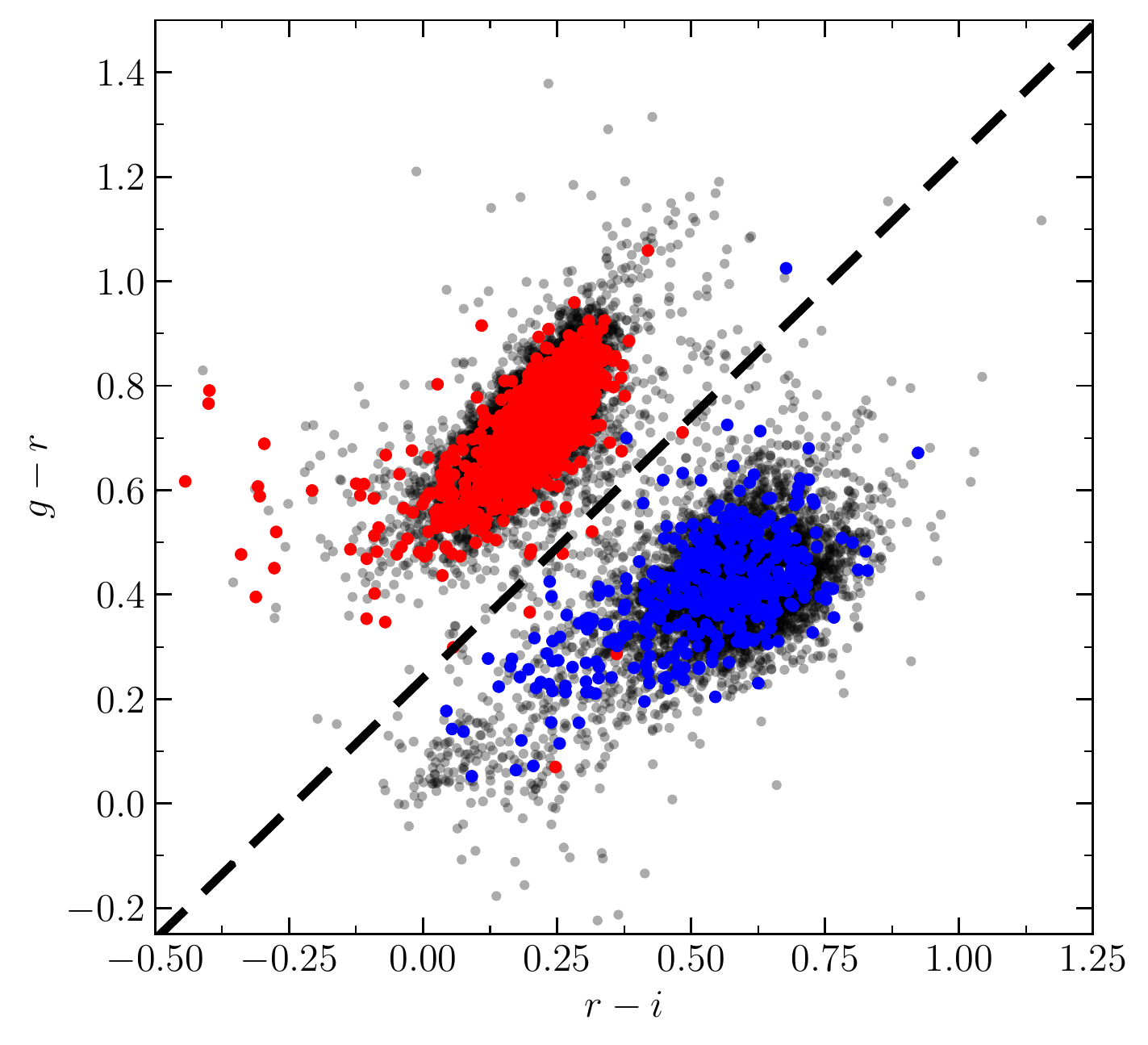}
\includegraphics[angle=0,width=\columnwidth]{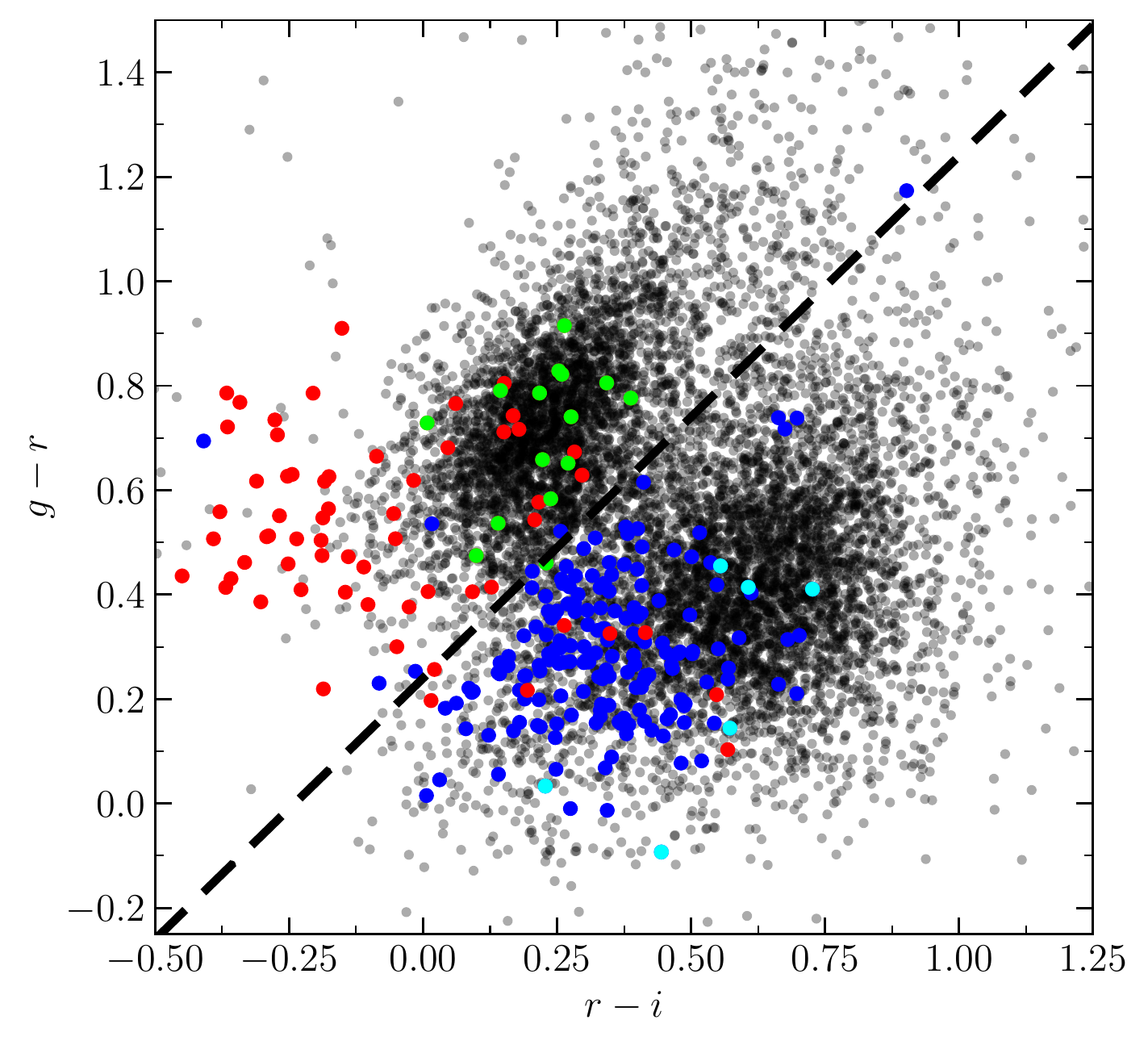}
\includegraphics[angle=0,width=\columnwidth]{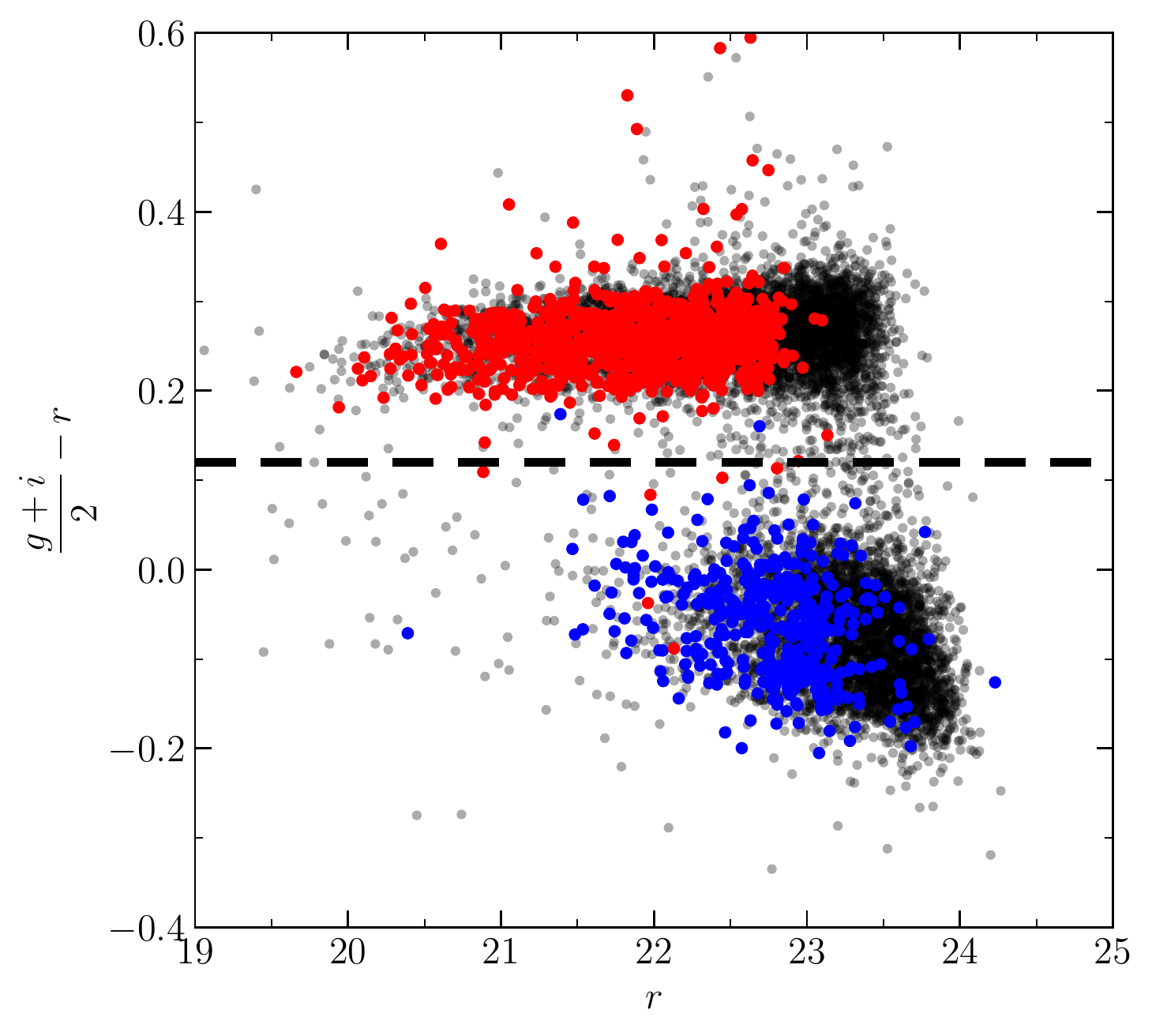}
\includegraphics[angle=0,width=\columnwidth]{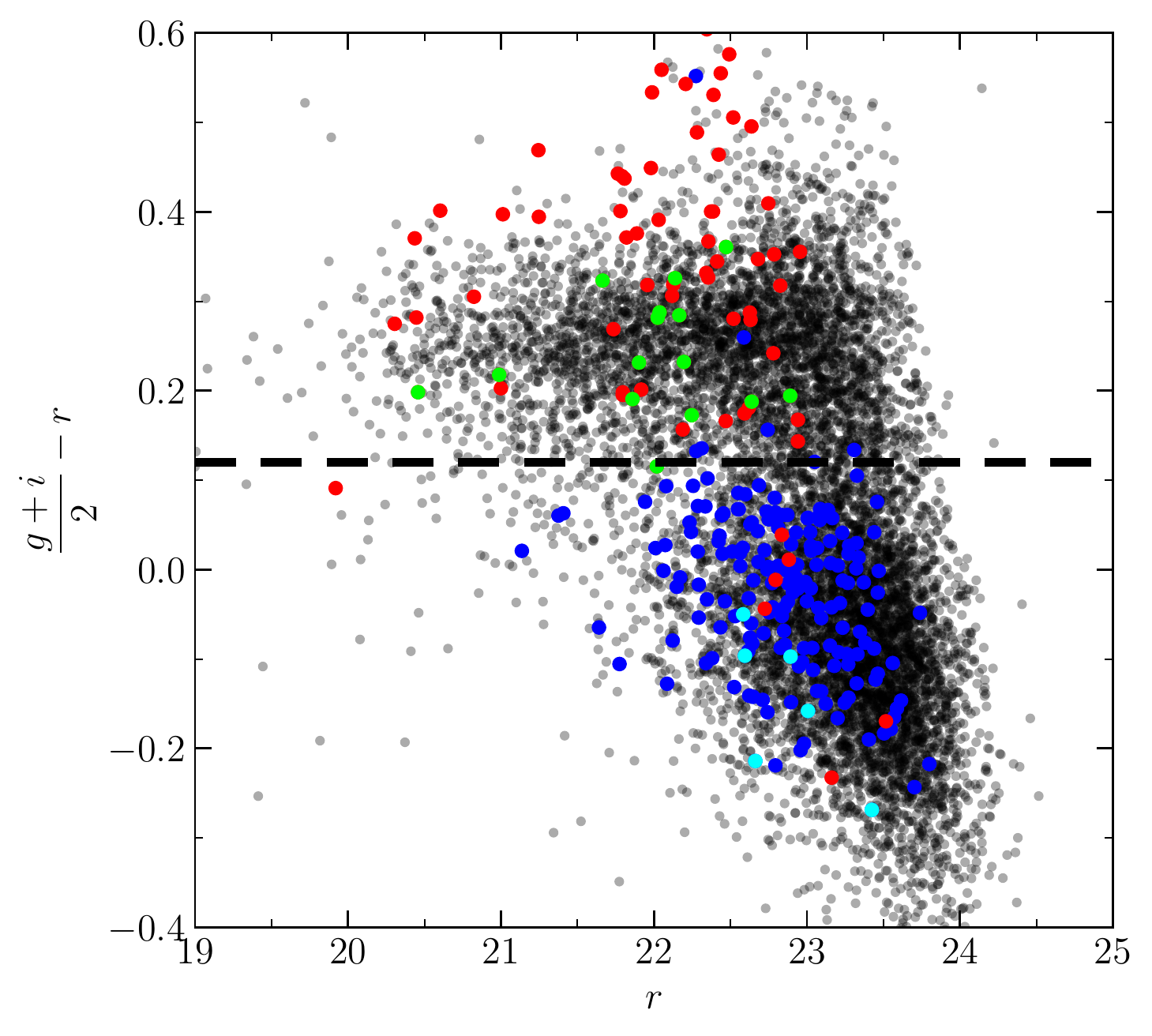}
\caption{\textit{Upper Left:} $g-r-i$ plotted for COSMOS. \textit{Upper Right:} $g-r-i$ plotted for HEROES. \textit{Lower Left:} $r$-excess vs. $r$ plotted for COSMOS. \textit{Lower Right:} $r$-excess vs. $r$ plotted for HEROES. In all panels, red points are spectroscopic HAEs, blue points are spectroscopic O3Es, and translucent black points are photometric candidates. The black dashed lines represent an $r$ excess of 0.12 magnitudes. This is the optimal separation criterion for the two populations in each field. In HEROES (right-hand panels), the DEIMOS identified HAEs are denoted by green points, and the DEIMOS identified O3Es are denoted by cyan points. Note that the sources targeted in HEROES with HYDRA (blue and red points) tend to be bluer than the DEIMOS sources. This phenomenon is due to the deliberate selection of apparently extreme sources in the early stages of this work for HYDRA observations. The DEIMOS sources---originally selected based on narrowband excesses---are better centered on the populations of photometric candidates.}
\label{fig:gri}
\end{figure*}

While the $i-z-Y$ broadband selection method is a reliable method to construct large samples of HAEs and O3Es in a large volume and to faint magnitudes, it fails to distinguish between the two. However, HAEs and O3Es can be separated using the $g-r-i$ color-color space, primarily due to  redshift dependent line contamination in the $r$-band (see below). We demonstrate this method for each field in the upper plots of Figure~\ref{fig:gri}, where we show $g-r$ versus $r-i$ for the sample of all photometric EELG candidates (black points), spectroscopically confirmed HAEs (red points), and spectroscopically confirmed O3Es (blue points). \cite{ly07} demonstrated a similar method, using $B-R_C-i'$ to distinguish HAEs from O3Es in NB921 (which overlaps the $z$ filter) excess sources (see \citealt{ly07},~Fig.~7).

In Figure \ref{fig:gri}, the populations of HAEs and O3Es are separated into two distinct regions, characterized primarily by the strength of the $r$-excess.  There are multiple explanations for this effect. At redshifts $0.3<z<0.42$ where the \Ha line falls in the $z$ filter, the [OIII] doublet happens to lie in $r$, exiting $r$ at redshift $z \approx 0.42$. Thus, as long as \Ha falls in $z$, a corresponding [OIII] doublet will fall in $r$ and will show an $r$ excess relative to $g$ and $i$. This separation might also be partially caused by continuum effects. Regardless of the underlying mechanism, the HAE population is cleanly observationally separated from the O3E sample in both fields. 

The lower plots of Figure \ref{fig:gri} further illustrate this effect, showing $r$ excess ($\frac{g+i}{2}-r$) versus $r$. It is clear that HAEs and O3Es can be distinguished from one another in 1410/1418 (99.4\%, COSMOS) and 244/261 (93.9\%, HEROES) of cases with an $r$ excess that HAEs will exceed and O3Es will fall below. In each of our fields, this $r$-excess value corresponded to 0.12 magnitudes. 

Combining these conditions with the initial $i-z-Y$ cuts yields the following color criteria for HAE and O3E broadband identification:

For HAEs at $0.3 \lesssim z \lesssim 0.42$: $\frac{g+i}{2}-r>0.12$, $i-z>0.17$, $z-y<0.00$, and $(i-z)+2(z-y)<0.30$.

And for O3Es at $0.7 \lesssim z \lesssim 0.86$: $\frac{g+i}{2}-r<0.12$, $i-z>0.17$, $z-y<0.00$, and $(i-z)+2(z-y)<0.30$.

These color cuts yield final photometric samples of 6211 HAEs and 4259 O3Es in COSMOS, and 39,109 HAEs and 52,276 O3Es in HEROES. From these photometric source counts, it is evident that the ratio of HAEs to O3Es is a factor of $\sim$2 higher in COMSOS than in HEROES. We attribute this discrepancy to cosmic variance (see also Section~\ref{sec:lfs}) and to differences in data quality and photometric calibration between the two datasets.

\section{Luminosity Functions}
\subsection{Lines Fluxes and Luminosities}
\label{sec:lineflux}
Using our samples of photometric HAEs and O3Es, we next construct luminosity functions (LFs) for each to test our selection in comparison to analogous narrowband studies. We use the broadband excess to calculate the line luminosity for each source---much like how a narrowband magnitude is used with an overlapping broadband measurement---to derive a line luminosity \citep[e.g.,][Eq. 3]{matthee2015}. In our broadband calculation, we estimate the continuum flux density in our central filter ($z$) by taking the average of the flux densities in the neighboring broadband filters ($i$ and $Y$). We then assume that the difference (excess) between the measured $z$ flux and the estimated continuum flux is due to the emission line of interest and other nearby emission lines. From this excess, we derive the emission line fluxes and luminosities using our spectroscopic redshifts and the filter transmission curve.

We show the full derivation of the expression for the \Ha luminosity in the Appendix. In brief, we consider the relative contributions of flux in $i$, $z$, and $Y$ from \Ha and [SII]$\lambda6716,6731$ as a function of redshift. Using our selected sample of zCOSMOS spectra, we determine median line flux ratios of [SII]$\lambda6716/$\Ha$=0.13$ and  [SII]$\lambda6731/$\Ha$=0.10$.  In the initial calculation, we neglect the [NII] doublet, as it is rarely detected with confidence in the zCOSMOS spectra, and instead correct for [NII] contamination in our final overall calibration (see below).
We then covert to a line luminosity using standard luminosity distance in our assumed cosmology. To account for the unknown redshifts of the purely photometric \Ha sources, we compute the average luminosity of each source over the redshift range $z=0.3-0.42$, taking into account the different luminosity distances and filter transmission efficiencies as a function of redshift. 

We compute the fluxes of the [OIII]$\lambda$5007,4959 and H$\beta$ lines in a similar manner. We assume a flux ratio of 3:1 in the [OIII] doublet, and we adopt [OIII]$\lambda$5007/H$\beta = 2.52$ based on the median of our selected sample of zCOSMOS spectra. We again show the full calculation in the Appendix.

We verify the accuracy of these line flux calculations using the selected zCOSMOS sample. We first flux calibrate the zCOSMOS spectra by integrating the spectra through the broadband filter bandpasses and renormalizing the spectra to match the fluxes measured in the HSC photometry. We then extract line fluxes from the spectra by fitting Gaussians to two groups of lines: 1) H$\beta$ and [OIII]$\lambda\lambda$4959,5007, and 2) \Ha, [NII]$\lambda\lambda$6548,6583, and [SII]$\lambda6716,6731$ (if available in the spectral range). We fit each group's lines simultaneously, using multiple Gaussians of the same width with fixed ratios of the line centers to account for any minor errors in the reported spectroscopic redshifts. We use the integrated areas of the fixed Gaussians as line flux measurements. We then compare these fitted line fluxes to our broadband derived line fluxes. 

For HAEs, the broadband method overestimates the line flux by 6.3\% relative to the median, so we rescale our broadband derived \Ha fluxes down by the same 6.3\% to compensate for the offset. We attribute this rescaling to contamination from the [NII] doublet, which is rarely detected with confidence in most spectra. After this rescaling, the distribution of errors ($\log(F_{\textrm{\Ha ,phot}})-\log(F_{\textrm{\Ha,spec}})$) is roughly symmetric with a standard deviation of 0.159 and shows no correlation with measured line luminosity. Thus, on average, our broadband derived \Ha line fluxes are accurate to within $\sim$37\%.

For O3Es, the broadband method underestimates the line flux by 29\% relative to the median, so we again apply an appropriate rescaling. We expect that this offset is due to some contamination of the continuum levels in the $i$ and Y bands by other emission lines, such as [OIII]$\lambda$4363 and H$\gamma$, as well as underlying continuum shape effects. The resulting distribution of errors ($\log(F_{\textrm{[OIII],phot}})-\log(F_{\textrm{[OIII],spec}})$) is also roughly symmetric and uncorrelated with measured line luminosity. The distribution has a standard deviation of 0.125, indicating an average [OIII]$\lambda5007$ flux accuracy of $\sim30\%$.

We attribute the scatter and initial offsets on these measurements to the inherent uncertainties in deriving line fluxes from a broadband excess, in which  parameters such as line ratios and continuum slopes/shapes cannot be determined. For instance, we calculate the assumed fixed line ratios from the median of the spectroscopic sample, which may not be accurate for an individual source. Despite this, due to the symmetries of the error distributions and the lack of correlation with measured line luminosity, these uncertainties can be largely ignored for populations of HAEs and O3Es, as the individual uncertainties will average out when considered in aggregate. 

It may also be possible to derive [OIII]$\lambda$5007 line luminosities at $z\sim0.3$ from the HAE samples using the $r$-band excess. However, this may be more challenging than deriving luminosities from the $z$-band excess, as the HSC $r$-filter is nearly twice as wide as the $z$-filter ($\sim$1400\AA{} vs $\sim$760\AA{}), thus limiting its sensitivity to line emission by a corresponding factor of $\sim$2 and making it more susceptible to continuum effects. We intend to investigate this in a future work.

\subsection{Equivalent Width Analysis}
\label{sec:ew}
We calculate observed-frame EWs for the samples by subtracting the calculated line luminosities (\Ha, [NII]$\lambda\lambda6548,6583$, and [SII]$\lambda\lambda6716,6731$ for HAEs, and [OIII]$\lambda\lambda4959,5007$ and H$\beta$ for O3Es) from the total $z$-filter flux to determine a local continuum level. We then divide the derived line luminosities by their corresponding continuum levels. We again verify the accuracy of these results by comparing to the fitted zCOSMOS lines fluxes and EWs. The distributions of errors on the estimated photometric EWs ($\log(\textrm{EW}_{\textrm{phot}})-\log(\textrm{EW}_{\textrm{spec}})$) are roughly symmetric and centered on zero for both HAEs and O3Es.

\begin{figure}[h]
\centering
\includegraphics[angle=0,width=\columnwidth]{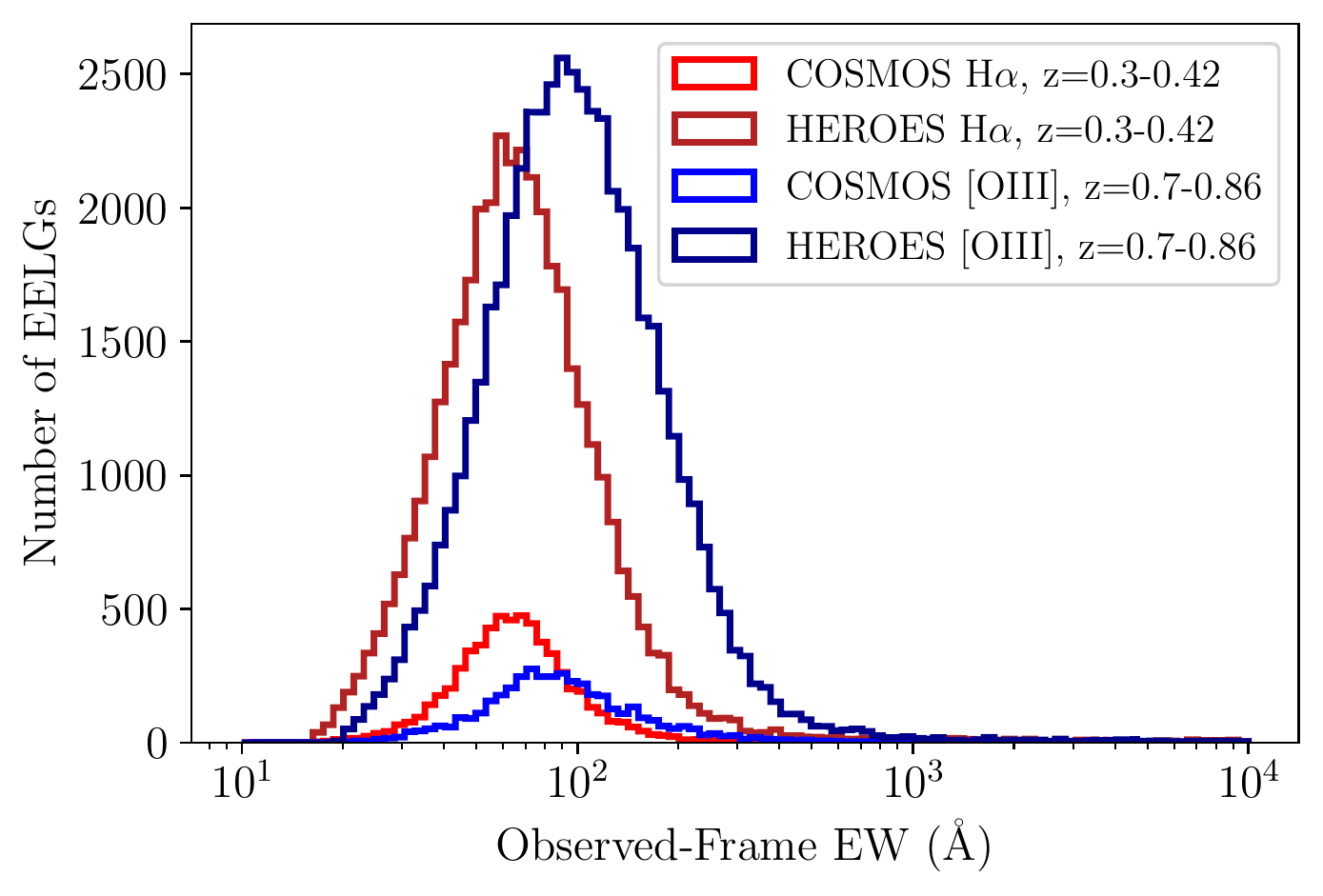}
\caption{Distributions of observed-frame EWs for HAEs and O3Es in COSMOS and HEROES. The histogram bins are equal width (0.05 $\log{\textrm{\AA}}$) in log space and are consistent between all four curves.}
\label{fig:ew}
\end{figure}

We show the distributions of observed-frame EWs for HAEs and O3Es in COSMOS and HEROES in Figure~\ref{fig:ew}. The medians for HAEs are 64~\AA{} and 65~\AA{} (rest-frame 47~\AA{} and 48~\AA{}), respectively. Both HAE distributions cut off at 16~\AA{}, the minimum observed-frame EW permitted by our color cuts, and have extended tails to high EW. At the very high EW end, many of these sources may be spurious; however, they may also be good candidates for studies of the most extreme star-forming galaxies or AGN. The O3E EW medians are 86~\AA{} and 96~\AA{} (rest-frame 48~\AA{} and 54~\AA{}), respectively. The O3E distributions have very similar shapes to the HAE distributions, with low end cutoffs at 19~\AA{} and extended high EW tails. These median EWs also show broad consistency with previous measurements of EWs with redshift \citep[e.g.,][their Figure 5]{labbe13}.

\subsection{Contamination Corrections}
As demonstrated in Figure \ref{fig:red_hist}, our spectroscopic samples are 94\% EELGs. In constructing LFs from the photometric samples, we corrected the samples to account for this $\sim6\%$ contamination fraction. We performed this correction as a function of line luminosity, calculating the fraction of spectroscopic sources that are both photometrically and spectroscopically identified as HAEs or O3Es for each luminosity bin and rescaling the photometric LF by this purity fraction. By calculating this fraction, we implicitly require spectroscopic coverage in each luminosity bin, thus ensuring reliable photometric luminosity calibration at all luminosities of interest.

\subsection{AGN Correction}
\begin{figure}[h]
\centering
\includegraphics[angle=0,width=\columnwidth]{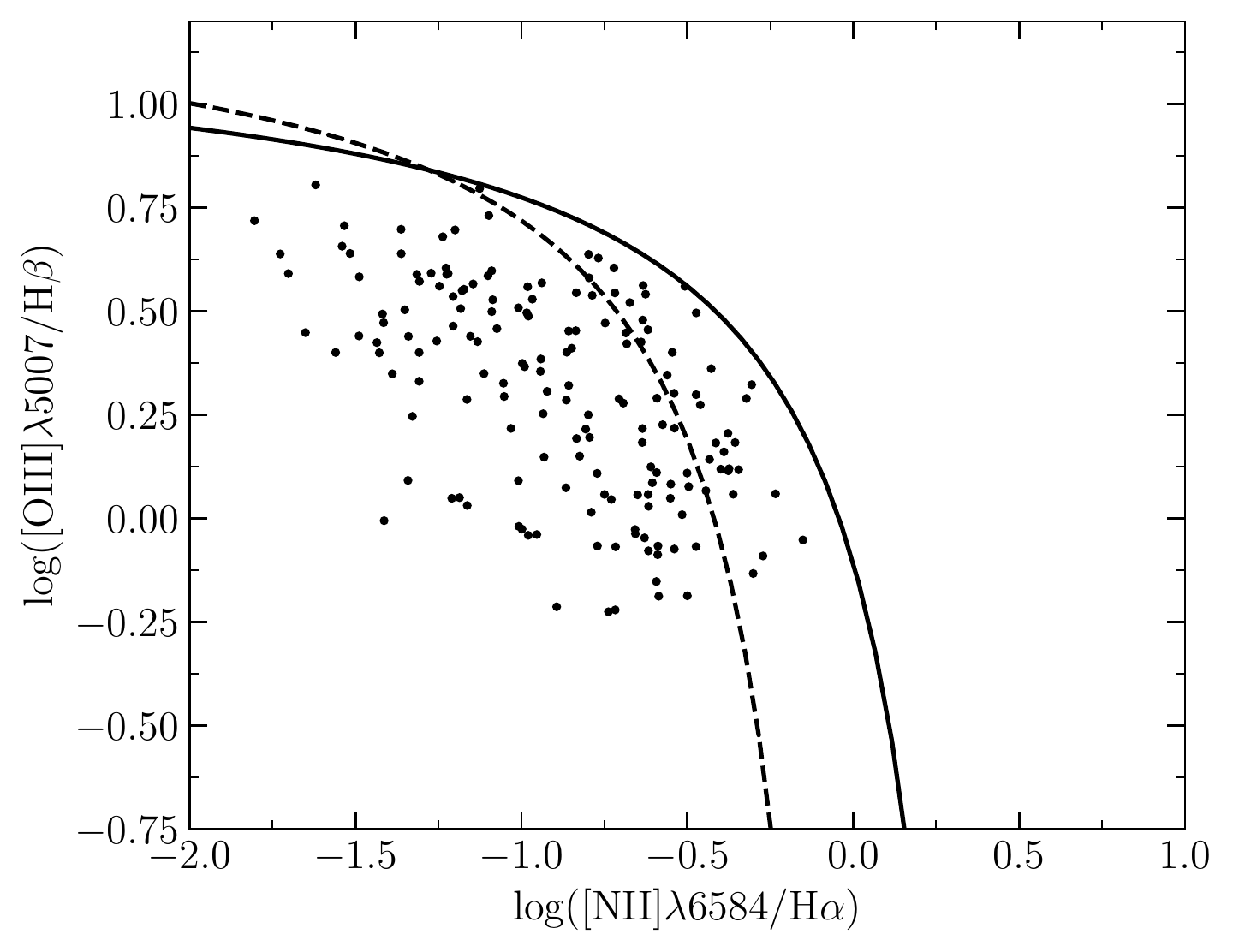}
\caption{BPT diagram constructed from the highest S/N subsample of the selected zCOSMOS sample. The dashed curve is the \cite{kauffmann03} empirical demarcation, and the solid black curve is the theoretical \cite{kewley01} line.}
\label{fig:BPT}
\end{figure}

We expect our $z$-excess samples to be contaminated by AGN. We attempted to correct for this contamination by producing a BPT diagram \citep{baldwin81} for the selected zCOSMOS sample, which we restricted to 232 sources that passed signal-to-noise cuts on the [NII] line (see Figure \ref{fig:BPT}). Despite consisting of only the highest S/N objects in the selected zCOSMOS sample, we found the star-forming track of the BPT diagram to be quite noisy and thus unreliable for differentiating star-forming galaxies from AGN. This noisy BPT diagram has been seen in previous studies and seems to be inherent in the zCOSMOS dataset \citep[e.g.,][their Figure 3]{bongiorno10}. The issue of AGN rejection via the BPT diagram is further complicated by the low signal-to-noise of the [NII]$\lambda$6584 line in fainter sources, and by the rarity of bright sources. 

In a similar \Ha study using a narrowband filter (CFHT/WIRcam LowOH2, center $\sim$11870 \AA{}), \cite{sobral13,sobral15Ha,sobral16} faced similar challenges with rejecting AGN, and they ultimately adopted constant AGN contamination fractions of 10\% at redshift $z=0.4$ and 15\% at redshift $z=0.84$ up to $L\sim L^*$, based on C-COSMOS X-ray data \citep{elvis09} and archival \textit{Spitzer}/IRAC data. Above $L^*$, they observed steep increases (AGN fraction $=0.38\log(L_{H\alpha})-15.8$) based on spectroscopic follow-up of their most luminous identified HAEs at $z=0.8$. 

We adopted their results in correcting our own \Ha LF, taking $L^*=10^{42}$~erg~s$^{-1}$ as the dividing value between the constant and linear corrections. 

\subsection{Completeness Correction}
We utilized Monte Carlo simulations to test the photometric completeness of our samples. We selected a star-forming galaxy template spectrum (\texttt{graz01\_00050.dat}) from the \texttt{EAZY} photometric redshift fitting code \citep{EAZY}, which was based on the \texttt{PEGASE.2} galaxy spectral synthesis code \citep{fioc99}. We removed the emission lines from the template to produce a continuum model. We then modeled the \Ha, H$\beta$, H$\gamma$, [OII]$\lambda$3727, [OIII]$\lambda$5007,4959,4363, and [NII]$\lambda$6583,6548 emission lines as 1D Gaussians and inserted them into the model spectrum, varying the redshift, \Ha luminosity, and \Ha EW. To determine the other line luminosities, we used the case B ratio for H$\beta$ and H$\gamma$, and the median ratios in the spectroscopic sample of line fluxes to the \Ha line fluxes for non-Hydrogen lines. To account for reddening, we calibrated and applied the \cite{calzetti00} extinction law to the rest-frame spectra  using an attenuation of \Ha of $A_{\Ha}=1.0$ magnitudes, as is commonly assumed in the literature \citep[e.g.,][]{hopkins04, takahashi07, sobral13, sobral15Ha, matthee17}. We chose the \cite{calzetti00} extinction law for its applicability to star-forming galaxies. Moreover, at rest-frame wavelengths $\lambda \gtrsim 3000$ \AA{}, most extinction curves (e.g., LMC, SMC, \citealt{calzetti00}) offer similar optical-NIR slopes such that differences would only be appreciable in the observed-frame $g$-filter \citep{salim20}.

We simulated HAE spectra for all permutations of $\log{(L_\textrm{\Ha})}=40.50$, 40.75, 42.00, ... 43.00~erg~s$^{-1}$, $z=0.300$, 0.325, 0.350, ... 0.425, and rest-frame EW$_{\Ha}=30, 40$, 50, 75, 100, 125, 150~\AA{}. We similarly simulated O3E spectra for all permutations of $\log(L_{\textrm{[OIII]}})=41.00$, 41.25, 41.50, ... 43.00~erg~s$^{-1}$, $z=0.700$, 0.725, 0.750, ... 0.875, and rest-frame EW$_{\textrm{[OIII]}}=30$, 40, 50, 75, 100, 125, 150~\AA{}. 

We produced artificial HAEs and O3Es from these model spectra as 2D Gaussian point sources with a FWHM of 0\farcs9, which is the median of the photometric dataset. We derived broadband magnitudes for these sources by integrating over the model spectra and $grizY$ filter response curves. We then converted the simulated broadband magnitudes into imaging counts based on the stacked imaging zeropoint and effective exposure time. We used these counts to set the amplitude of the 2D Gaussian source models. For each model spectrum, we inserted artificial sources in the processed HEROES and COSMOS imaging (in $g,r,i,z,Y$), corresponding to a simulated source density of $\sim1$~arcmin$^{-2}$, which is sufficient to well sample the imaging with minimal source-source overlap. We used \texttt{sep} \citep{sep}, a Python derivative of \texttt{SExtractor} \citep{sextractor}, to detect sources in the $z$-filter with a 5$\sigma$ detection threshold and to measure forced aperture magnitudes for the detected sources in $g,r,i,z,Y$. We then applied our photometric selection cuts to the resulting source catalog and determined what fraction of the simulated sources we recovered.  We took this fraction as our completeness value for the combination of the given redshift, \Ha or [OIII] EW, and \Ha or [OIII] luminosity. 

\begin{figure}[h]
\centering
\includegraphics[angle=0,width=\columnwidth]{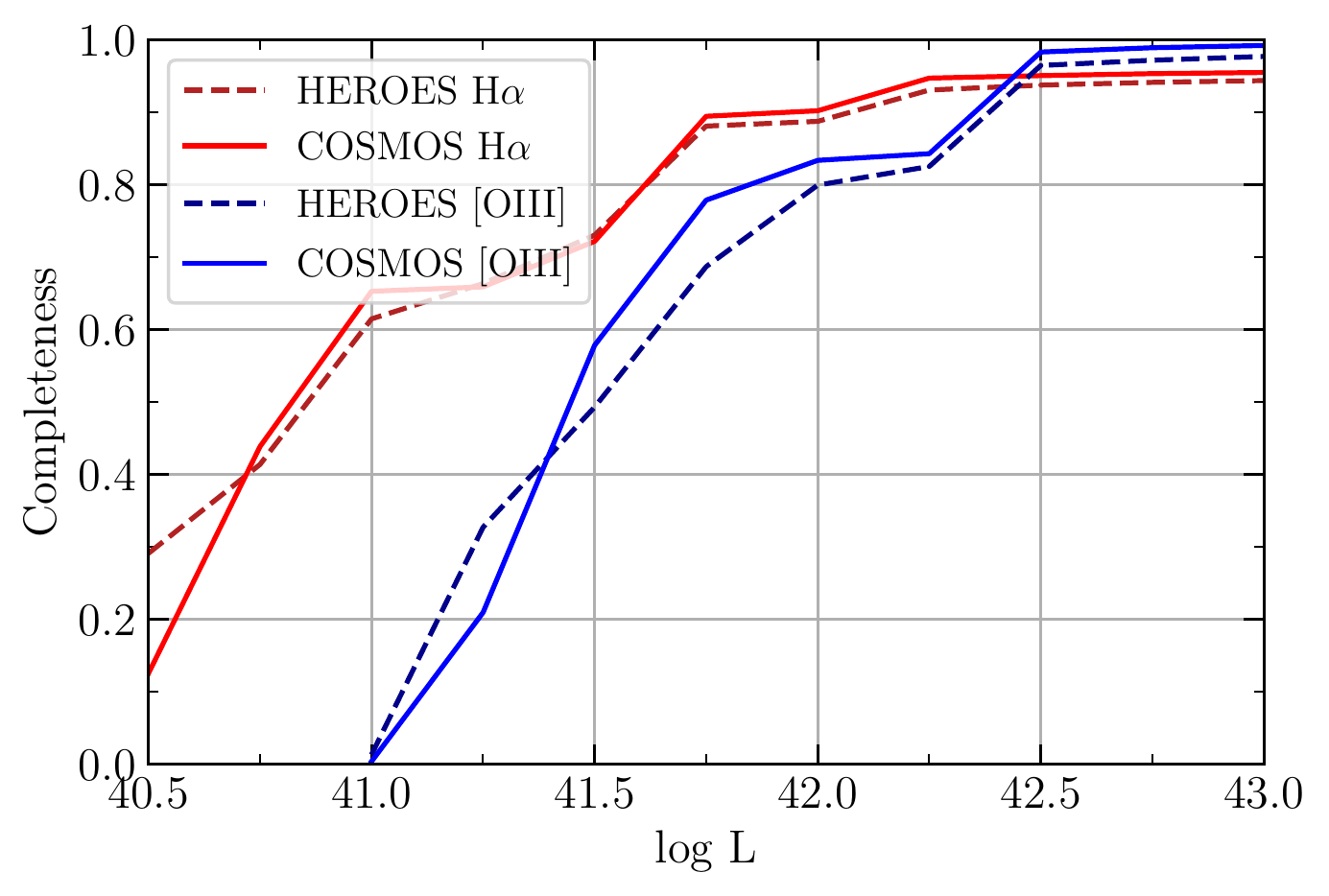}
\caption{Average completeness of the photometric HAE (red) and O3E (blue) samples for both HEROES (dashed) and COSMOS (solid) fields.} 
\label{fig:completeness}
\end{figure}

In Figure~\ref{fig:completeness} we show the completeness averaged over the chosen redshift ranges (HAEs: $0.30<z<0.42$, O3Es: $0.70<z<0.86$) and over the rest-frame EWs seen in the selected zCOSMOS spectral catalog as a function of line luminosity. As expected, the COSMOS field's superior imaging quality offers marginally higher completeness at the bright end for both species.  We assume a $\pm10\%$ error on our completeness at all luminosities to account for any minor effects from our choice of spectral template, simulated line ratios, and dust attenuation law. 

\subsection{Luminosity Functions}
\label{sec:lfs}
\begin{figure}[h]
\centering
\includegraphics[angle=0,width=\columnwidth]{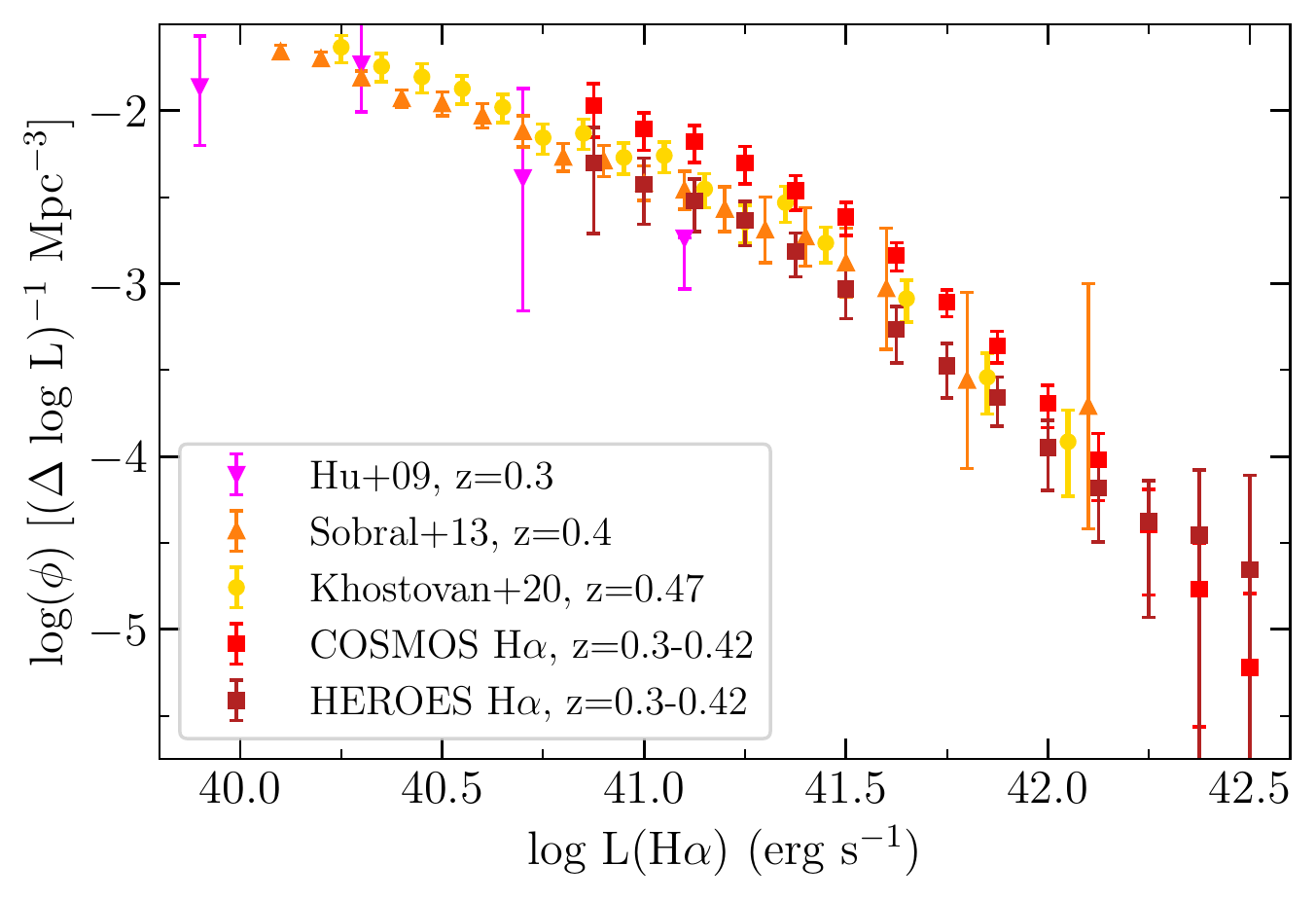}
\caption{Incompleteness and AGN corrected photometric LFs for \Ha at $0.30 < z < 0.42$ (COSMOS: red squares, HEROES: dark red squares) and literature samples (see figure legend). Note that none of the LFs shown correct for dust attenuation (the dust attenuation adjustment of 1 magnitude has been removed from the \citealt{sobral13} dataset). The \cite{hu09} dataset has been scaled up by a factor of 25 to account for their limited selection of USELs \citep[see][Figure 14 for details]{hu09}. We provide the data in this figure in Table \ref{tab:LF}.} 
\label{fig:HaLF}
\end{figure}

\begin{figure}[h]
\centering
\includegraphics[angle=0,width=\columnwidth]{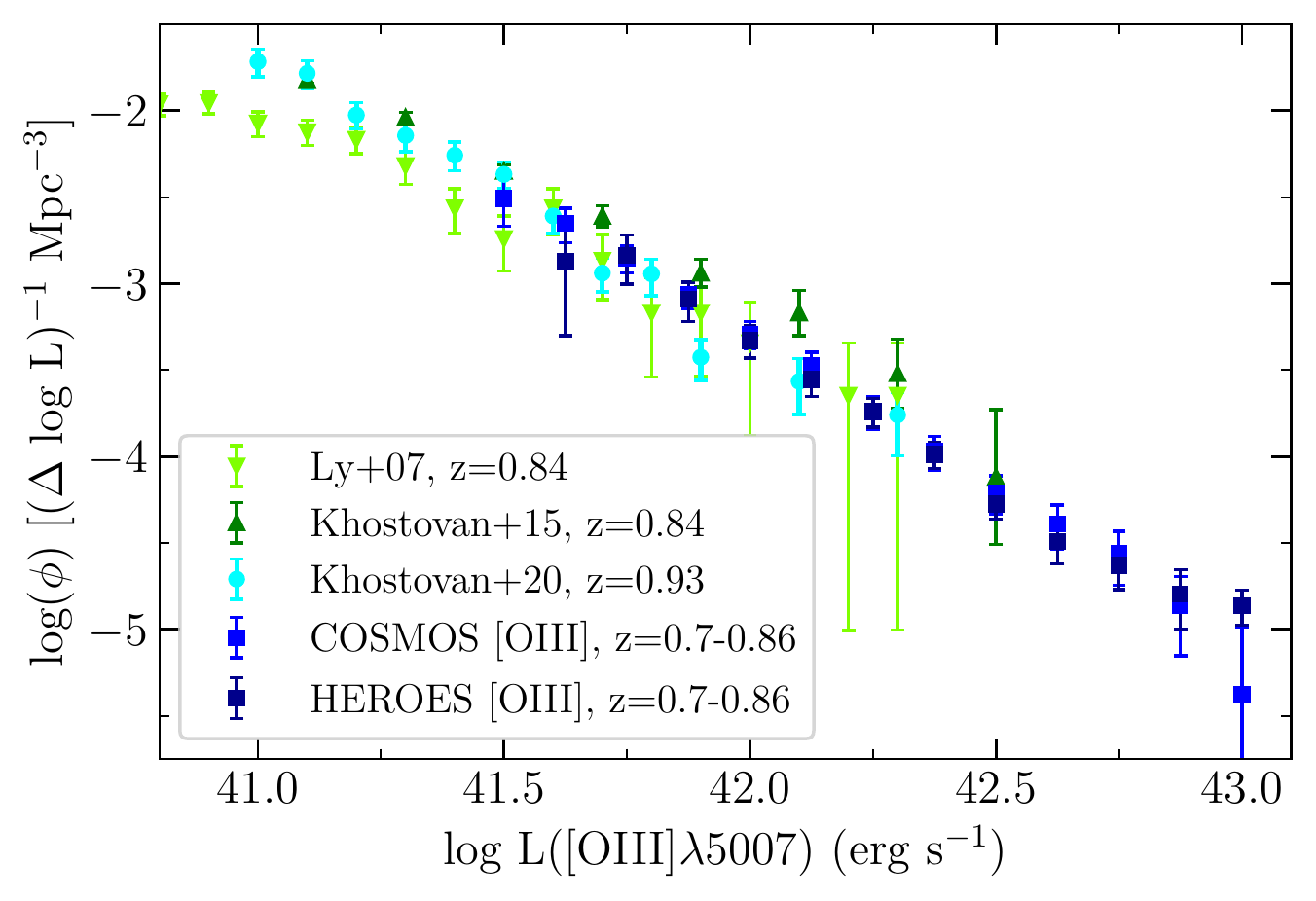}
\caption{Incompleteness corrected photometric LFs for [OIII] at $0.70 < z < 0.86$ (COSMOS: blue squares, HEROES: dark blue squares) and literature samples (see figure legend). Note that none of the LFs shown are corrected for dust attenuation and that the \cite{khostovan15} data are for a combined [OIII]+H$\beta$ LF. We provide the data in this figure in Table \ref{tab:LF}.} 
\label{fig:OIIILF}
\end{figure}

Figures \ref{fig:HaLF} and \ref{fig:OIIILF} show the photometric \Ha and [OIII] LFs. Our comoving volumes are calculated from the 3.15/34.2 deg$^2$ observed areas (COSMOS/HEROES), and the redshift ranges $0.30<z<0.42$ and $0.70 < z < 0.86$. The COSMOS field has a comoving volume of $8.15\times 10^5$~Mpc$^3$ at $0.30 < z < 0.42$, and $3.17\times 10^6$~Mpc$^3$ at $0.70 < z < 0.86$. HEROES encompasses $8.85\times 10^6$~Mpc$^3$ at $0.30 < z < 0.42$, and $3.44\times 10^7$~Mpc$^3$ at $0.70 < z < 0.86$. 

As described above, we correct our \Ha LFs for completeness (via the completeness simulations), for sample purity (using the fraction of positive spectroscopic confirmations at a given luminosity), and for AGN contamination \citep[using the prescription from][]{sobral15Ha}. We cut off our LFs at the faint end, ignoring luminosity bins with $<50\%$ completeness. We characterize our uncertainties assuming simple Poissonian statistics ($\sigma_x=\sqrt{x}$) for the populations of photometric candidates, spectroscopically confirmed HAEs or O3Es, and spectroscopically blank/indeterminate objects for each luminosity bin. We propagate these errors, along with an assumed 10 percentage point error on the completeness correction, to produce our plotted 1$\sigma$ error bars. As a result, our LF uncertainties are a function of both the number of photometric sources in a bin (which scales with survey area) and the spectroscopic completeness of each sample. Thus, due to its much higher spectroscopic completeness (13.8\% verses 0.39\%), our COSMOS \Ha LF has smaller uncertainties than HEROES \Ha LF, despite HEROES' $\sim10$x larger comoving volume. 

Up to $\log{(L_\textrm{\Ha})}=42.0$~erg~s$^{-1}$, the COSMOS \Ha LF exhibits a $\sim0.3$~dex offset from the HEROES \Ha LF. We attribute this offset to the aforementioned differences in data quality between the two datasets, any photometric calibration offsets between the two datasets, and to cosmic variance. Using the methodology of \cite{driver10}, we estimate cosmic variance in our surveyed \Ha-sensitive COSMOS and HEROES volumes of 20\% and 10\%, respectively. While these values alone are insufficient to make up the $\sim0.3$~dex offset, when combined with dataset to dataset variance, and considering the agreement with literature samples (see below), we are unconcerned by this offset. 

We compare our \Ha LFs to literature samples from \cite{hu09} (pink downward triangles), \citealt{sobral13} (orange upward triangles), and \citealt{LAGER20} (yellow circles). All of these samples were selected from narrowband imaging surveys (NB816 and NB921; NB921 and NB964, respectively) and are therefore more sensitive to line emission than our broadband selection, while probing correspondingly smaller comoving volumes due to the inherently narrow passbands. Given the $\sim$800~\AA{} width of the $z$-filter compared to the $\sim$120~\AA{} width of the narrowband filters, our selection probes $\sim$6x the comoving volume of a narrowband study for an equivalent survey area at the cost of a corresponding $\sim$6x increase in the uncertainty on the redshifts of photometric candidates. 

It is important to note that none of the plotted LFs are corrected for dust attenuation. \cite{sobral13} assumed (as is common in the literature; see also \citealt{LAGER20}, Section 3.4 and references therein) a 1 magnitude attenuation of the \Ha line flux. We remove this correction in Figure \ref{fig:HaLF} by offsetting the \cite{sobral13} dataset by $-0.4$ dex from their reported luminosity values. While we considered performing a dust attenuation correction of our own data, we found that our spectroscopic datasets were insufficiently calibrated to provide reliable H$\alpha/$H$\beta$ line ratios for calibrating dust attenuation with the Balmer decrement. \cite{LAGER20} noted the same difficulty with the zCOSMOS dataset and instead opted for a dust correction based on $g-r$ colors. We refrained from using such a correction to avoid any potential bias introduced by using $g-r$ color for both dust attenuation and differentiation between HAEs and O3Es. 

\cite{hu09} reported that their USEL LF offered agreement with the more general \Ha LF of \cite{tresse98} if the general LF were multiplied by 0.04. We thus scale up the \cite{hu09} data by a factor of 25 to compensate for this effect and to provide a direct comparison to the other studies.

With these adjustments applied, we find that the HEROES \Ha LF is within the $1\sigma$ error bounds of the other studies. The offset on the COSMOS \Ha LF marginally separates it from the other studies, but when cosmic variance and the different redshift intervals are taken into consideration, this LF is still in agreement with the literature samples. From these results, we conclude that our 5-broadband $z$-excess selection technique is reliable and well-calibrated for selecting HAEs at $0.30<z<0.42$. 

We construct our [OIII] LFs in much the same way as the \Ha LFs, except that we do not include an AGN contamination correction. This allows us to make a more direct comparison with other studies that also lack an AGN correction, such as \cite{LAGER20}. In contrast to the \Ha LFs, our COSMOS and HEROES [OIII] LFs show remarkable agreement with one another, with a median offset of 0.02 dex. Both LFs show strong agreement within the error bounds of the narrowband selected literature samples from \cite{ly07} (light green downward pointing triangles) and \cite{LAGER20} (cyan circles). The data from \cite{khostovan15} (dark green upward pointing triangles) are a combined [OIII]+H$\beta$ LF, which may explain the slight offset from the other studies around $\log{(L_{\textrm{[OIII]}})}=42.0$~erg~s$^{-1}$.  Due to the lack of AGN corrections, all of the studies deviate from a Schechter-like shape above $\log{(L_{\textrm{[OIII]}})}\sim41.75$~erg~s$^{-1}$. Due to the large comoving volumes sampled by our broadband selection, our LFs are able to push to $\log{(L_{\textrm{[OIII]}})}=43.0$~erg~s$^{-1}$, brighter than any of the literature studies. At this bright end, the LFs strongly resemble a power law and are likely dominated by contributions from AGN. We intend to investigate the AGN contribution at $z\sim0.7$ in a future study. Due to our decreased line sensitivity when compared to narrowband surveys and our corresponding $Y<23$ constraint to ensure strong photometric detections, we do not probe as faint as the narrowband literature studies, but we nonetheless demonstrate the reliability of our broadband selection method. 

This result is particularly exciting in the context of \textit{VRO}. Over its 10 year main survey, \textit{VRO} expects to reach $5\sigma$ imaging depths of 25.8, 27.0, 27.2, 27.0, 25.7, and 24.4 for $ugrizy$, respectively, over $\sim$18,000~deg$^2$ \citep{LSSTSciBook}. These depths are all equal to or deeper than our requirements of $g$,$r$,$i$,$z<25$, and $y<23$ and will likely meet our requirements by the end of the first survey year (assuming uniform coverage in all filters and efficient data processing). Using our selection cuts, and scaling from the number density of sources in the NEP, we expect that \textit{VRO} may detect $\sim$21 million HAEs at $0.30<z<0.42$ and $\sim$28 million O3Es at $0.70<z<0.86$. Thus, \textit{VRO} EELG studies may very easily constrain the bright end of the LFs, subject to suitable AGN rejection. As with our COSMOS and HEROES samples, the \textit{VRO} data will need to be carefully calibrated with spectroscopic measurements of sources across its survey field to both inform its exact color and magnitude cuts and constrain contamination levels in photometric EELG samples. Whether analyzed as a contiguous $\sim$18,000~deg$^2$ or more practically in smaller segments, \textit{VRO} will reveal an unprecedented number of HAEs and O3Es with the $z$-excess method.

\begin{deluxetable*}{ccccc}[ht]
\renewcommand\baselinestretch{1.0}
\tablewidth{0pt}
\tablecaption{Luminosity Function Data}
\label{tab:LF}
\tablehead{$\log{L}$ & COSMOS $\log{\Phi_{\textrm{\Ha}}}$ & HEROES $\log{\Phi_{\textrm{\Ha}}}$ & COSMOS $\log{\Phi_{\textrm{[OIII]}}}$ & HEROES $\log{\Phi_{\textrm{[OIII]}}}$ \cr (erg s$^{-1}$) & \multicolumn{4}{c}{($\Delta\log{L}^{-1}$ Mpc$^{-3}$)}}
\startdata
40.875 & -1.971$^{+0.128}_{-0.183}$ & -2.304$^{+0.206}_{-0.407}$ & \nodata & \nodata  \\
41.000 & -2.107$^{+0.094}_{-0.121}$ & -2.425$^{+0.150}_{-0.231}$ & \nodata & \nodata  \\
41.125 & -2.180$^{+0.094}_{-0.120}$ & -2.520$^{+0.126}_{-0.178}$ & \nodata & \nodata  \\
41.250 & -2.302$^{+0.095}_{-0.121}$ & -2.635$^{+0.109}_{-0.147}$ & \nodata & \nodata  \\
41.375 & -2.463$^{+0.089}_{-0.112}$ & -2.815$^{+0.108}_{-0.144}$ & \nodata & \nodata  \\
41.500 & -2.616$^{+0.085}_{-0.106}$ & -3.032$^{+0.123}_{-0.172}$ & -2.507$^{+0.116}_{-0.160}$ & \nodata \\
41.625 & -2.838$^{+0.074}_{-0.090}$ & -3.265$^{+0.133}_{-0.193}$ & -2.652$^{+0.089}_{-0.111}$ & -2.874$^{+0.211}_{-0.427}$ \\
41.750 & -3.107$^{+0.071}_{-0.085}$ & -3.475$^{+0.130}_{-0.186}$ & -2.854$^{+0.071}_{-0.085}$ & -2.838$^{+0.119}_{-0.164}$ \\
41.875 & -3.359$^{+0.082}_{-0.101}$ & -3.660$^{+0.119}_{-0.165}$ & -3.064$^{+0.070}_{-0.084}$ & -3.091$^{+0.100}_{-0.129}$ \\
42.000 & -3.694$^{+0.105}_{-0.139}$ & -3.949$^{+0.156}_{-0.247}$ & -3.292$^{+0.071}_{-0.085}$ & -3.329$^{+0.084}_{-0.103}$ \\
42.125 & -4.020$^{+0.152}_{-0.236}$ & -4.181$^{+0.180}_{-0.314}$ & -3.473$^{+0.076}_{-0.092}$ & -3.554$^{+0.082}_{-0.100}$ \\
42.250 & -4.396$^{+0.206}_{-0.404}$ & -4.376$^{+0.236}_{-0.556}$ & -3.737$^{+0.085}_{-0.106}$ & -3.74$^{+0.075}_{-0.090}$ \\
42.375 & -4.769$^{+0.265}_{-0.796}$ & -4.455$^{+0.378}_{-\infty}$ & -3.969$^{+0.086}_{-0.108}$ & -3.988$^{+0.068}_{-0.081}$ \\
42.500 & -5.221$^{+0.428}_{-\infty}$ & -4.654$^{+0.545}_{-\infty}$ & -4.209$^{+0.096}_{-0.124}$ & -4.274$^{+0.074}_{-0.089}$ \\
42.625 & \nodata & \nodata & -4.389$^{+0.109}_{-0.145}$ & -4.493$^{+0.099}_{-0.128}$ \\
42.750 & \nodata & \nodata & -4.562$^{+0.128}_{-0.183}$ & -4.629$^{+0.106}_{-0.141}$ \\
42.875 & \nodata & \nodata & -4.865$^{+0.172}_{-0.288}$ & -4.796$^{+0.139}_{-0.206}$ \\
43.000 & \nodata & \nodata & -5.376$^{+0.390}_{-\infty}$ & -4.862$^{+0.090}_{-0.114}$ \\
\enddata
\end{deluxetable*}

\section{Conclusions}

The main results of our work are as follows:
\begin{enumerate}
\item{We introduced a novel $z$-excess 5-filter broadband selection technique for identifying HAEs at $0.3<z<0.42$ and O3Es at $0.7<z<0.86$.}
\item{Using 3.15~deg$^2$ of HSC-SSP broadband data in the COSMOS field in conjunction with archival spectroscopic data from zCOSMOS and DEIMOS 10K, as well as new observations from WIYN/HYDRA, we tested this selection technique.}
\item{We expanded and applied the technique to 34.2~deg$^2$ of HEROES broadband data in the NEP field and new spectroscopy from Keck/DEIMOS and WIYN/HYDRA.}
\item{We presented spectroscopic catalogs of HAEs and O3Es identified in our WIYN/HYDRA observations (see appendix).}
\item{We introduced analytical expressions for estimating \Ha or [OIII] line flux from broadband magnitudes, and we calibrated the expressions using the spectroscopic datasets.}
\item{Using our broadband selection technique and calibrated line flux expressions, we constructed \Ha and [OIII] LFs and found strong agreement with narrowband literature studies.}
\end{enumerate}

Based on the 94\% fidelity in selecting spectroscopic HAEs and O3Es, the 98.5\% accuracy in differentiating them from each other, and the strong agreement with literature LFs, we conclude that the $z$-excess 5-filter broadband selection technique presented in this work is both accurate and effective in identifying EELGs at $z<1$ for either spectroscopic follow-up in small fields, or for population analysis in large fields, without the need for costly full spectroscopic samples. With the ever increasing sizes of current and future multi-band photometric surveys, such as will be obtained with the \textit{VRO}, this technique may be key in identifying and characterizing unprecedented numbers of EELGs in a straightforward manner.

\acknowledgements
We thank the anonymous referee for an insightful report that helped us to improve this work. We gratefully acknowledge support for this research from Jeff and Judy Diermeier through a Diermeier Fellowship (B.E.R.), a Wisconsin Space Grant Consortium Graduate and Professional Research Fellowship (A.J.T.), a Sigma Xi Grant in Aid of Research (A.J.T.), NSF grants AST-1716093 (E.M.H., A.S.) and AST-1715145 (A.J.B), the trustees of the William F. Vilas Estate (A.J.B.), and the University of Wisconsin-Madison, Office of the Vice Chancellor for Research and Graduate Education with funding from the Wisconsin Alumni Research Foundation (A.J.B.). 

This paper is based in part on data collected from the Subaru Telescope. The Hyper Suprime-Cam (HSC) collaboration includes the astronomical communities of Japan and Taiwan, and Princeton University. The HSC instrumentation and software were developed by the National Astronomical Observatory of Japan (NAOJ), the Kavli Institute for the Physics and Mathematics of the Universe (Kavli IPMU), the University of Tokyo, the High Energy Accelerator Research Organization (KEK), the Academia Sinica Institute for Astronomy and Astrophysics in Taiwan (ASIAA), and Princeton University. Funding was contributed by the FIRST program from Japanese Cabinet Office, the Ministry of Education, Culture, Sports, Science and Technology (MEXT), the Japan Society for the Promotion of Science (JSPS), Japan Science and Technology Agency (JST), the Toray Science Foundation, NAOJ, Kavli IPMU, KEK, ASIAA, and Princeton University. 

This paper also makes use of data collected at the Subaru Telescope and retrieved from the HSC data archive system, which is operated by Subaru Telescope and Astronomy Data Center at National Astronomical Observatory of Japan. Data analysis was in part carried out with the cooperation of Center for Computational Astrophysics, National Astronomical Observatory of Japan.

This paper is based in part on data collected from the Keck~II Telescope. The W.~M.~Keck Observatory is operated as a scientific partnership among the 
California Institute of Technology, the University of California, and NASA, and was made possible by the generous financial support of the W.~M.~Keck Foundation. 

This work is based in part on observations at Kitt Peak National Observatory, NSF's National Optical-Infrared Astronomy Research Laboratory, which is operated by the Association of Universities for Research in Astronomy (AURA) under a cooperative agreement with the National Science Foundation. The WIYN Observatory is a joint facility of the NSF's National Optical-Infrared Astronomy Research Laboratory, Indiana University, the University of Wisconsin-Madison, Pennsylvania State University, the University of Missouri, the University of California-Irvine, and Purdue University.

This material is based upon work supported by NASA under Award No. RFP20\_9.0 issued through Wisconsin Space Grant Consortium. Any opinions, findings, and conclusions or recommendations expressed in this material are those of the authors and do not necessarily reflect the views of the National Aeronautics and Space Administration.

This research made use of \textit{Astropy}, a community-developed core Python package for Astronomy \citep{astropy:2013, astropy:2018}.

The authors wish to recognize and acknowledge the important cultural role and reverence that the summit of Maunakea has always had within the  Hawaiian community. We are fortunate to have the opportunity to conduct observations from this site.


\bibliography{ref1}

\appendix{}
\label{append:a}

Here we show the derivation of an estimated \Ha line luminosity from $i$, $z$, and $Y$ broadband fluxes, corresponding broadband filter transmission curves, and an assumed redshift (see Section \ref{sec:lineflux}).
\newline\newline
\noindent Assume that the continuum flux density in the z filter, $z_c$, is such that:

\begin{equation}  z_c=(i_c+y_c)/2 \end{equation}

\noindent In the detected redshift range, \Ha must be in the $z$-filter to cause a $z$-excess detection, but the [SII] doublet may fall in the neighboring $Y$-filter. 
\noindent The flux density in each filter is given by

\begin{equation}  i=i_c \,, \end{equation}

\begin{equation}  z=z_c + \frac{\epsilon_{z\Ha} \Ha + \epsilon_{z6716} [\textrm{SII}]_{6716} + \epsilon_{z6731} [\textrm{SII}]_{6731}}{\Delta \lambda_z}\,, \end{equation}

\begin{equation}  \textrm{and }y=y_c + \frac{\epsilon_{y6716} [\textrm{SII}]_{6716} + \epsilon_{y6731} [\textrm{SII}]_{6731}}{\Delta \lambda_y} \end{equation}

\noindent Here, $x_c$ is the continuum flux density in filter $x$, $\Delta \lambda_x$ is the effective width of filter $x$, $\Ha,[\textrm{SII}]_{6716},[\textrm{SII}]_{6731}$ are the integrated observed-frame line fluxes of $\Ha,[\textrm{SII}]_{6716},[\textrm{SII}]_{6731}$, respectively, and $\epsilon_{xn}$ is the filter efficiency of filter $x$ at wavelength $n$.

We next assume a constant line flux ratios of $[\textrm{SII}]_{6716}/\Ha=0.13$ and $[\textrm{SII}]_{6731}/\Ha=0.10$. Inserting this substitution and simplifying gives

\begin{equation}  z=z_c + (\epsilon_{z\Ha} + 0.13\epsilon_{z6716} + 0.10\epsilon_{z6731}) \frac{\Ha}{\Delta \lambda_z}\,, \end{equation}

\begin{equation}  y=y_c + (0.13\epsilon_{y6716} + 0.10\epsilon_{y6731})\frac{\Ha}{\Delta \lambda_y} \end{equation}

\noindent Multiplying (1) by a factor of two and plugging (2, 5, and 6) into (1) gives

\begin{equation}  2z-2 (\epsilon_{z\Ha} + 0.13\epsilon_{z6716} + 0.10\epsilon_{z6731}) \frac{\Ha}{\Delta \lambda_z}= i + y - (0.13\epsilon_{y6716} + 0.10\epsilon_{y6731}) \frac{\Ha}{\Delta \lambda_y}  \end{equation}

\noindent Collecting like terms gives

\begin{equation}   i+y-2z = \Ha \left[\frac{0.13\epsilon_{y6716} + 0.10\epsilon_{y6731}}{\Delta \lambda_y} - \frac{2 (\epsilon_{z\Ha} + 0.13\epsilon_{z6716} + 0.10\epsilon_{z6731})}{\Delta \lambda_z}  \right]   \end{equation}

\begin{equation}   \Ha= \frac{i+y-2z}{\frac{0.13\epsilon_{y6716} + 0.10\epsilon_{y6731}}{\Delta \lambda_y} - \frac{2 (\epsilon_{z\Ha} + 0.13\epsilon_{z6716} + 0.10\epsilon_{z6731})}{\Delta \lambda_z}}  \end{equation}

\noindent With $d_L$ as the redshift determined luminosity distance of the $[\textrm{OIII}]_{5007}$ source, the rest-frame $[\textrm{OIII}]_{5007}$ luminosity ($L$) is given by

\begin{equation}  \boxed{L = 4 \pi d_L^2 \Ha =\frac{4 \pi d_L^2 (i+y-2z)}{\frac{0.13\epsilon_{y6716} + 0.10\epsilon_{y6731}}{\Delta \lambda_y} - \frac{2 (\epsilon_{z\Ha} + 0.13\epsilon_{z6716} + 0.10\epsilon_{z6731})}{\Delta \lambda_z}}} \end{equation}

\noindent For the special case in which all three lines fall within $z$, $\epsilon_y=0$ and $\epsilon_i=0$ for all lines, greatly simplifying the expression to

\begin{equation}  \boxed{L = 4 \pi d_L^2 \Ha =\frac{4 \pi d_L^2 (z-\frac{i+y}{2})\Delta \lambda_z}{(\epsilon_{z\Ha} + 0.13\epsilon_{z6716} + 0.10\epsilon_{z6731})}} \end{equation}

Here we show the derivation of an estimated [OIII]$\lambda$5007 line luminosity from $i$, $z$, and $Y$ broadband fluxes, corresponding broadband filter transmission curves, and an assumed redshift (see Section \ref{sec:lineflux}).
\newline\newline
\noindent Assume that the continuum flux density in the z filter, $z_c$, is such that:

\begin{equation}  z_c=(i_c+y_c)/2 \end{equation}

\noindent In the detected redshift range $[\textrm{OIII}]_{5007}$ must fall in the $z$-filter to cause a $z$-excess detection, but H$\beta$ and/or $[\textrm{OIII}]_{4959}$ may fall in the neighboring $i$-filter. This can mitigated by modifying the redshift range to require that all three lines fall in $z$ (see simplified special case at the end). 
\noindent The flux density in each filter is given by

\begin{equation}  i=i_c + \frac{\epsilon_{iH\beta} H\beta + \epsilon_{i4959} [\textrm{OIII}]_{4959}}{\Delta \lambda_i}\,, \end{equation}

\begin{equation}  z=z_c + \frac{\epsilon_{zH\beta} H\beta + \epsilon_{z4959} [\textrm{OIII}]_{4959} + \epsilon_{z5007} [\textrm{OIII}]_{5007}}{\Delta \lambda_z}\,, \end{equation}

\begin{equation}  \textrm{and }y=y_c \end{equation}

\noindent Here, $x_c$ is the continuum flux density in filter $x$, $\Delta \lambda_x$ is the effective width of filter $x$, $H\beta,[\textrm{OIII}]_{4959},[\textrm{OIII}]_{5007}$ are the integrated observed-frame line fluxes of $H\beta,[\textrm{OIII}]_{4959},[\textrm{OIII}]_{5007}$, respectively, and $\epsilon_{xn}$ is the filter efficiency of filter $x$ at wavelength $n$.

We next assume constant line flux ratios of $[\textrm{OIII}]_{4959}/[\textrm{OIII}]_{5007}=0.33$ and $H\beta/[\textrm{OIII}]_{5007}=0.40$ in order to have all line fluxes in terms of $[\textrm{OIII}]_{5007}$. Inserting this substitution and simplifying gives

\begin{equation}  i=i_c + (0.40\epsilon_{iH\beta}  + 0.33\epsilon_{i4959})\frac{[\textrm{OIII}]_{5007}}{\Delta \lambda_i} \,,\end{equation}

\begin{equation}  z=z_c + (0.40\epsilon_{zH\beta} + 0.33\epsilon_{z4959} + \epsilon_{z5007})\frac{[\textrm{OIII}]_{5007}}{\Delta \lambda_z} \end{equation}

\noindent Multiplying (12) by a factor of two and plugging (15, 16, and 17) into (12) gives

\begin{equation}  2z-2 (0.40\epsilon_{zH\beta} + 0.33\epsilon_{z4959} + \epsilon_{z5007})\frac{[\textrm{OIII}]_{5007}}{\Delta \lambda_z}= i-(0.40\epsilon_{iH\beta}  + 0.33\epsilon_{i4959})\frac{[\textrm{OIII}]_{5007}}{\Delta \lambda_i} + y \end{equation}

\noindent Collecting like terms gives

\begin{equation}   i+y-2z = [\textrm{OIII}]_{5007}\left[\frac{0.40\epsilon_{iH\beta}  + 0.33\epsilon_{i4959}}{ \Delta \lambda_i} -2\frac{0.40\epsilon_{zH\beta}/ + 0.33\epsilon_{z4959} + \epsilon_{z5007}}{\Delta \lambda_z}\right]   \end{equation}

\begin{equation}   [\textrm{OIII}]_{5007}= \frac{ i+y-2z}{\frac{0.40\epsilon_{iH\beta}  + 0.33\epsilon_{i4959}}{ \Delta \lambda_i}  + -\frac{2(0.40\epsilon_{zH\beta} + 0.33\epsilon_{z4959} + \epsilon_{z5007})}{\Delta \lambda_z}}  \end{equation}

\noindent With $d_L$ as the redshift determined luminosity distance of the $[\textrm{OIII}]_{5007}$ source, the rest-frame $[\textrm{OIII}]_{5007}$ luminosity ($L$) is given by

\begin{equation}  \boxed{L = 4 \pi d_L^2 [\textrm{OIII}]_{5007} =\frac{4 \pi d_L^2 (i+y-2z)}{\frac{0.40\epsilon_{iH\beta}  + 0.33\epsilon_{i4959}}{ \Delta \lambda_i} -\frac{2(0.40\epsilon_{zH\beta} + 0.33\epsilon_{z4959} + \epsilon_{z5007})}{\Delta \lambda_z}}} \end{equation}

\noindent For the special case in which all three lines fall within $z$, $\epsilon_i=0$ for all lines, greatly simplifying the expression to

\begin{equation}  \boxed{L = 4 \pi d_L^2 [\textrm{OIII}]_{5007} = 4 \pi d_L^2\frac{z-\frac{i+y}{2}}{0.40\epsilon_{zH\beta} + 0.33\epsilon_{z4959} + \epsilon_{z5007}} \Delta \lambda_z} \end{equation}

\end{document}